\documentclass[10pt,a4paper]{article}
\usepackage[a4paper, left=2.0cm, right=2.0cm, top=2cm]{geometry}
\usepackage[normalem]{ulem}
\usepackage[utf8]{inputenc}
\usepackage{amsmath, amsthm, amssymb, amsfonts}     
\usepackage{mathrsfs}                               
\usepackage{accents}                                
\usepackage{natbib}                                 
\usepackage{hyperref}                               
\usepackage{doi}                                    
\usepackage{authblk}                                
\usepackage{siunitx}                                
\usepackage{booktabs}                               
\usepackage{multirow}                               
\usepackage{microtype}                              
\usepackage{caption}
\usepackage{subcaption}                             
\usepackage{graphicx}

\let\savetabular\tabular
\def\tabular{\footnotesize\baselineskip=12pt\savetabular}

\title{Two-Phase Flow Simulations of Surface Waves in Wind-Forced Conditions}
\author[1]{Malte Loft\thanks{malte.loft@tuhh.de} }
\author[1]{Niklas K\"uhl}
\author[2]{Marc P.\ Buckley}
\author[2]{Jeffrey R.\ Carpenter}
\author[3]{Michael Hinze}
\author[4]{Fabrice Veron}
\author[1]{Thomas Rung}

\affil[1]{Institute for Fluid Dynamics and Ship Theory, Hamburg University of Technology, Am Schwarzenberg-Campus 4, D-21075 Hamburg, Germany}
\affil[2]{Institute of Coastal Ocean Dynamics, Helmholtz-Zentrum Hereon,  Max-Planck-Straße 1, D-21502 Geesthacht, Germany}
\affil[3]{Modeling, Simulation \& Optimization of Complex Systems, Universität Koblenz-Landau, Universitätsstraße 1, D-56070 Koblenz, Germany}
\affil[4]{School of Marine Science and Policy, University of Delaware, Newark, DE 19716, USA}

\begin{document}

\maketitle


\begin{abstract}
The paper is devoted to two-phase flow simulations and investigates the ability of a diffusive interface Cahn-Hilliard Volume-of-Fluid model to capture the dynamics of the air-sea interface at geophysically relevant Reynolds numbers. It employs a hybrid filtered/averaging Improved Detached Eddy Simulation method to model turbulence, and utilizes a continuum model to account for surface tension if the diffuse interface is under-resolved by the grid. A numerical wind-wave tank is introduced to limit computational costs and results obtained for two wind-wave conditions are analyzed in comparison to experimental data at matched Reynolds numbers. The focus of the comparison is on both time-averaged and wave-coherent quantities, and includes pressure, velocity as well as modeled and resolved Reynolds stresses. In general, numerical predictions agree well with the experimental measurements and reproduce many wave-dependent flow features. Reynolds stresses near the water surface are found to be especially important in modulating the critical layer height. It is concluded that the diffusive interface approach proves to be a promising method for future studies of air-sea interface dynamics in geophysically relevant flows. 
\end{abstract}

\section{Introduction}
\label{sec:introduction}
The momentum and mechanical energy fluxes across the air-sea interface are an essential component of the global energy budget. It has been estimated that over 80\% of the kinetic energy within the ocean originates from the mechanical action of the wind upon the wavy ocean surface \citep{wunsch2004vertical}. These fluxes are controlled by the small-scale processes within the coupled atmospheric and oceanic wave boundary layers \citep{sullivan2010dynamics}. Global circulation models are not able to resolve these small-scale processes, and rely on parameterizations of the wind stress. The parameterizations quantify both pressure drag and viscous stress contributions, whose influence depends on the specific wind-wave conditions. Moreover, the pathways leading to the partitioning of wind energy into wave growth, wave breaking, and current generation remain poorly understood \citep{sullivan2010dynamics}. At the same time, the influence of individual wave field parameters on the airflow dynamics is also unexplained. This is due, in part, to the technical challenges involved in observing and simulating the physics in the fairly thin, dynamic two-phase flow regime of the wavy boundary layer, rendering detailed experiments or numerical simulations challenging. 

From a numerical viewpoint, previous investigations were mostly restricted to single-phase approaches, where the air phase was predominantly studied. Such aerodynamic simulations specify a given mean free surface evolution, e.g., a plane progressive surface wave train. The (mean) free surface behavior is either obtained from precursor hydrodynamic simulations \citep{sullivan2018turbulent}, (semi-)analytical descriptions based on empirically confirmed statistics \citep{sullivan2014large}, or simpler analytical expressions, e.g., Stokes waves as discussed in \citet{yang2017direct}. Results of these ''one-way'' couplings between a prescribed water phase and a simulated air phase show that the relation between the sea state and the wind stresses remain uncertain in high wind conditions, where nonlinear turbulent processes are dominant \citep{edson2013exchange}. Moreover, frequently employed interface descriptions, such as roughness-supported law-of-the-wall expressions, must be adjusted to wave parameters -- for example, to the wave steepness or the wave age -- to alter the roughness of the free surface and thereby the surface stress \citep{husain2019boundary}. However, despite comprehensive parameterizations of the atmospheric boundary layer and the related drag coefficient, the physical processes acting in the atmospheric and oceanic wavy boundary layers are not well known \citep{sullivan2010dynamics}.

Single-phase aerodynamic simulations benefit from superior efficiency and an inherently sharp interface. However, the benefits come at the expense of an assumed sea state and hypothesized turbulent interface physics. This limits the air-sea flux simulation capabilities and motivates a two-phase flow framework. Mesh-based computational approaches for immiscible two-phase flows mainly refer to interface-capturing methods which reconstruct the free surface position from an indicator function. Within the class of interface-capturing methods, the Volume-of-Fluid (VoF) approach of \citet{hirt1981volume} is predominantly used. The method is conservative and capable of predicting (incipient) wave breaking and related influences on the generation of turbulence, vorticity, and ocean current, for example, as discussed by \citet{perlin2013breaking}. Breaking waves were previously simulated but with less attention to wind-wave interaction. Examples are found in \citet{iafrati2009numerical,iafrati2011energy} 
or \citet{lubin2015numerical}, who presented the first Large Eddy Simulation (LES) results obtained by a VoF approach for a plunging breaking wave problem. The latter study was confined to the generation of dissipative vortical structures under breaking waves. Employing a similar VoF method, \citet{hao2018simulation} reported Direct Numerical Simulation (DNS) results for wind over breaking waves scenarios. Most previous two-phase flow simulations did neglect capillary effects, which might be of interest for wind-wave interaction simulations, as indicated by the pronounced sensitivity to the roughness model outlined in \citet{husain2019boundary}. An exception is the work of \citet{deike2015capillary} and \citet{wu2022}. In the latter, similar wave parameters as in this study were investigated with a VoF method. Due to the Reynolds number limitations imposed by DNS, the investigated Reynolds numbers $\text{Re}_\lambda \approx 200$ are, however, significantly smaller than in this paper which reaches up to $\text{Re}_\lambda \approx 20,000$, cf.\ Table \ref{tab:scenarios}. 

An alternative to the VoF approach is the diffuse phase field model suggested by \citet{cahn1958free}. In contrast to VoF methods, the Cahn-Hilliard (CH) approach models the phase separation of immiscible fluids and naturally incorporates surface tension/capillary effects, which is why a related empirical model, such as the popular Continuum Method suggested by \citet{brackbill1992continuum}, is not necessarily required for interface resolving simulations. There exists a variety of CH models for two-phase flows. Examples refer to fluids with matched densities described in \citet{hohenberg1977theory}, identical viscosities outlined by \citet{jacqmin1999calculation}, or thermodynamically consistent approaches \citep[cf.][]{abels2012thermodynamically,garcke2019diffuse,lowengrub1998quasi}. The present research aims at simulations of the air-sea interface along a route outlined in \citet{eden2019energy}, i.e., using a CH-VoF approach published in \citet{kuhl2021cahn}. The model bridges VoF and CH concepts and can be used for different levels of interface resolution, i.e., in combination with implicit (CH-based) surface tension models in the resolved case, or explicit surface tension models in an under-resolved case. The two-phase approach is implemented into an implicit, multi-phase Finite-Volume (FV) method using unstructured grids and a free surface adapted Detached Eddy Simulation framework \citep[DES;][]{spalart2006new} to model turbulence. 

The paper is structured as follows: Sec.\ \ref{sec:compModel} introduces the computational model, including the building blocks specific to the two-phase flow treatment. Subsequently, the triple decomposition, originally suggested in \citet{phillips1957generation}, utilized to process the computed spatio-temporal results is described in  Sec.\ \ref{sec:data}. The fourth section is devoted to a simple verification study. Applications for two different experimentally reported wind-wave scenarios are presented in Sec.\ \ref{sec:Application}. Results include instantaneous fields, wave coherent and averaged quantities as well as turbulent stresses in comparison with experimental data. Final conclusions and future directions are outlined in Sec.\,\ref{sec:Conclusion}. Within the publication, Einstein summation convention is used for lower-case Latin subscripts. Vectors and tensors are defined with reference to Cartesian coordinates. 
\clearpage
\section{Computational Model}
\label{sec:compModel}
The computational model consists of four essential building blocks. The baseline FV procedure is briefly discussed in Sec.\ \ref{sec:numerics}. The  employed two-phase flow model and the hybrid filtered/averaged turbulence closure are outlined in Sec.\ \ref{Sec:two-phase} and \ref{sec:turb}. A numerical wind-wave tank, which aims to limit the computational cost, is introduced in Sec.\ \ref{sec:windWaveTankModel}. 

\subsection{Numerical Procedure} 
\label{sec:numerics}
The numerical procedure utilizes a FV approximation dedicated to Single Instruction Multiple Data (SIMD) implementations on a distributed-memory parallel CPU machine. Algorithms employed by the in-house procedure FresCo$^+$ are described in \citet{rung2009challenges},  \citet{yakubov2013hybrid} and \citet{kuhl2021adjoint}. They ground on the integral form of a generic Eulerian transport equation, with residuum $\mathrm{R}^{\varphi}$ for a scalar field $\varphi(x_i, t)$ exposed to the influence of a possibly non-linear source  $\mathcal{S}^\varphi$ in addition to a modeled (non-linear) gradient diffusion term and its diffusivity $\Gamma^{\mathrm{eff}}$ in a control volume $V$ bounded by the surface $S(V)$, viz. 
\begin{align}
    \label{equ:momentum_numerics}
    \int_V \mathrm{R}^\varphi = 0 \, , \qquad \to \qquad 
    \int_V \left[ \frac{\partial \varphi}{\partial t} - \mathcal{S}^\varphi \right] \mathrm{d} V + \oint_{S(V)}  \left[  u_{i} \varphi - \Gamma^{\mathrm{eff}} \frac{\partial \varphi}{\partial x_{i}} \right] n_i\,\mathrm{d}S = 0 \; . 
\end{align}
Here, $x_i$ refers to the Cartesian spatial coordinates, $u_i$ denotes to the Cartesian components of the velocity vector, $n_i$ features the face-normal vector and $t$ is the time. The sequential procedure uses the strong conservation form and employs a cell-centered, co-located storage arrangement for all transport properties. Spatial discretization assumes unstructured grids based on arbitrary polyhedral cells, which connect to a face-based data structure. Various turbulence-closure models are available with respect to (w.r.t.) statistical (RANS) or scale-resolving (LES, DES) approaches. The numerical integration refers to the mid-point rule. Diffusive fluxes are determined from second-order central differencing, and convective fluxes employ higher-order upwind biased interpolation formulae. An exception refers to the convection of the concentration field (cf.\ Sec.\ \ref{Sec:two-phase}), which employs the compressive High Resolution Interface-Capturing (HRIC) scheme published by \citet{muzaferija1998computation} as well as the Inter Gamma Differencing Scheme (IGDS) discussed by \citet{jasak1995interface} in conjunction with VoF simulations. In contrast, a computationally more robust first-order upwind scheme is used in conjunction with the CH-VoF method advocated in this paper. Time integration follows from a second-order accurate Implicit Three-Time Level (ITTL) scheme. Preconditioned Krylov-subspace solvers are used to solve the algebraic equation systems, and the global flow field is iterated to convergence using a pressure-correction scheme (SIMPLE). The procedure is parallelized using a domain decomposition method and the Message Passing Interface (MPI) communication protocol.

\subsection{Two-Phase Flow Model}
\label{Sec:two-phase}
The computational model assumes two immiscible, inert fluids  ($a$, $b$) featuring constant bulk densities ($\rho_\mathrm{a}$, $\rho_\mathrm{b}$) [kg/m$^3$] and bulk viscosities ($\mu_\mathrm{a}$, $\mu_\mathrm{b}$) [N$\cdot$ s/m$^2$]. Fluid $a$ is referred to as foreground fluid, i.e., the air phase, and fluid $b$ as background fluid, i.e., the water phase. Both fluids are assumed to share the kinematic field along the route of the VoF-approach suggested by \citet{hirt1981volume}. An Eulerian concentration field describes the spatial distribution of the fluids, where $c = c_\mathrm{a} = V_\mathrm{air}/V$ denotes the volume concentration of the foreground fluid, and the volume fraction occupied by the background fluid refers to $c_\mathrm{b} = V_\mathrm{sea} / V = ( V - V_\mathrm{air})/V = (1 - c)$.

\subsubsection{Concentration Transport}
\label{sec:conctrans}
Since the material properties of immiscible, inert fluids are invariable when using the VoF-method, the concentration is governed by a simple Lagrangian transport equation, i.e., $\mathrm{d} c_\mathrm{a}/ \mathrm{d} t  \, (=- \mathrm{d} c_\mathrm{b}/ \mathrm{d} t) = \mathrm{d} c/\mathrm{d} t =0$, which is translated into an Eulerian formulation before its discretization. On the contrary, diffuse interface methods replace the sharp interface with a thin layer where the fluids exchange mass fluxes. Cahn-Hilliard approaches can be separated into mass and volume conservative strategies and essentially augment the Lagrangian concentration transport by a velocity-divergence term and a non-linear, diffusive right-hand side (RHS) of order four, that  vanishes outside the interface region, cf.\ \citet{ding2007diffuse} and \citet{kuhl2021cahn}, viz. 
\begin{align}
    \frac{\mathrm{d} c}{\mathrm{d} t} = 
    \frac{\partial}{\partial x_k}
    \left[ M \frac{\partial \psi}{\partial x_k} 
    \right] - c \frac{\partial u_k}{\partial x_k} 
    \quad \to \quad
    \frac{\partial c}{\partial t} + \frac{\partial \, u_k c}{\partial x_k} =
    \frac{\partial}{\partial x_k}
    \left[ M \frac{\partial \psi}{\partial x_k} 
    \right]. 
    \label{equ:CHceq}
\end{align}
Here, $\psi(c, \partial^2 c/\partial x_k^2)$ [N/m$^2$] is a chemical potential, and $M(c)$ refers to a mobility parameter of dimension m$^4$/(N s). Following \citet{kuhl2021cahn}, the present study employs a mass conservative strategy together with an appropriate choice of $M$ and a frequently used ''double-well potential'', which yields
\begin{align}
     \psi &= 2 C_\mathrm{1} \left[ (2c^3 -3c^2+c) - 0.5 \left(\frac{C_\mathrm{2}}{C_\mathrm{1}}\right) \frac{\partial^2 c}{\partial x_k^2} \right]\; ,   
    \\
    \frac{\partial \psi}{\partial x_k}  & = 2 C_\mathrm{1} \left[ (6c^2 -6c+1) \frac{\partial c}{\partial x_k}  - 0.5 \left(\frac{C_\mathrm{2}}{C_\mathrm{1}}\right) \frac{\partial^3 c}{\partial x_k^3} \right].
    \label{equ:CHceqB}
\end{align}
Two parameters are involved in the definition of the chemical potential $\psi$. These read  $C_1  = \sigma/\gamma$ and  $C_2  = \sigma \, \gamma$, and both employ the surface tension $\sigma$ [N/m] and the interface thickness $\gamma$ [m]. The ratio $C_\mathrm{2}/C_\mathrm{1} \sim \gamma^2$ scales with the square of the interface thickness. Moreover, the product $C_\mathrm{1} \, M \sim \nu_\mathrm{c}$ [m$^2$/s] in (\ref{equ:CHceq}) describes a nonlinear apparent viscosity $\nu_\mathrm{c} = 2 C_\mathrm{1} \cdot M (6c^2 -6c+1)$. Evaluating the last term of $\nabla_k \psi$ in (\ref{equ:CHceqB}) requires sufficient grid resolution, i.e., the term can be neglected when the interface is under-resolved, which is the case in the present air-sea simulations. Mind that $\nu_\mathrm{c}$ vanishes at $c=(0.5 \pm \sqrt{3}/6)$  and is negative over approximately $58$\,\% of the inner transition regime between these roots, where it supports the phase separation process. 

Though the non-zero RHS of (\ref{equ:CHceq}) appears to increase the complexity, it is beneficial for various reasons \citep{kuhl2021cahn}. On the one hand, it facilitates more sound interface physics, e.g., naturally includes surface tension effects when the grid resolution adequately resolves the phase transition. On the other hand, the use of robust/stability-preserving, upwind-biased convective approximations is supported. The present applications are devoted to under-resolved studies. This allows to cover larger spatio-temporal domains at higher Reynolds numbers but requires surface tension effects to be modeled by an auxiliary model, e.g., as described in Eqn.\ (\ref{equ:brackbill}). 
 
\subsubsection{Equation of State}
\label{sec:EoS}
An Equation of State (EoS) is used to extract the local flow properties from the concentration field, the bulk properties and a non-dimensional function $m(c)$, viz.
\begin{align}
    \rho = m^\mathrm{\rho}  \rho^\mathrm{\Delta} + \rho_\mathrm{b} 
    \qquad \qquad \mathrm{and} \qquad \qquad
    \mu = m^\mathrm{\mu}  \mu^\mathrm{\Delta} + \mu_\mathrm{b} \, , \label{equ:mater_prope}
\end{align}
where $\rho^\mathrm{\Delta} = \rho_\mathrm{a} - \rho_\mathrm{b}$, $\mu^\mathrm{\Delta} = \mu_\mathrm{a} - \mu_\mathrm{b}$ mark the respective bulk property differences. Although this is not necessary, the paper assigns $m^\mathrm{\mu} = m^\mathrm{\rho}$. Provisions on the EoS considered in this study aim to exclude non-physical, unbounded density states by restricting $m \in \left[0,1\right]$ and to recover the single-phase limit states, i.e.\ $m(c=1[0]) = 1[0]$, as discussed in \citet{kuhl2021cahn}. The simplest conceivable EoS $m^{(1)}$ corresponds to a bounded linear interpolation between the limit states and is usually employed by  VoF methods. More advanced alternatives $m^{(2)}$ and $m^{(3)}$  follow a hyperbolic tangent rule or its linearized version and employ a user-specified non-dimensional transition parameter $\gamma^\mathrm{m}$, viz. 
\begin{equation}
    \label{equ:eos_linear}
    m^\mathrm{(1)} = \begin{cases}
    0 &\text{if} \ c < 0 \\
    1 &\text{if} \ c > 1 \\
    c &\text{otherwise}
    \end{cases}
\end{equation}
as well as 
\begin{equation} 
    m^\mathrm{(2)} = \frac{1}{2} \left[ \tanh \left( \frac{2 c - 1}{\gamma^\mathrm{m}}\right) +1 \right] 
    \qquad \mathrm{or} \qquad 
	m^\mathrm{(3)} = \left\{  \begin{matrix}
	0 & \text{if } c < 0.5   (1 - \gamma^\mathrm{m})  \\
    1 & \text{if } c >0.5   (1 + \gamma^\mathrm{m})  \\
	\frac{2c+ 1 - \gamma^\mathrm{m}}{2\gamma^{\mathrm{m}}} & \text{otherwise.}
	\end{matrix} \right. 
    \label{equ:eos_hyperbol}
\end{equation}
The hyperbolic EoS complies only asymptotically with the limit states. Hence, an upper bound for the transition parameter reads $\gamma^\mathrm{m} \le 0.5$ to limit the error w.r.t.\ the limit states below $0.1$\,\%. Mind that $m^{\mathrm{(3)}}$ refers to a linearization of $m^{(2)}$ and is preferred in the present study due to the non-asymptotic characteristics.

In combination with a CH formulation, an hyperbolic EoS offers a decisive advantage for constructing a two-phase flow model that does not resolve the extremely thin interface and closely resembles the traditional VoF framework. Introducing the EoS (\ref{equ:mater_prope}) into the mass conservative continuity equation yields an expression for the divergence of the velocity field, viz. 
\begin{align}
    \frac{\partial \rho}{\partial t} + \frac{\partial \, u_k \rho}{\partial x_k} = 0
    \qquad  \to \qquad 
    \frac{\partial u_k}{\partial x_k} =
     \frac{\dot\rho_\mathrm{a}}{\rho_\mathrm{a}}+ \frac{\dot \rho_\mathrm{b}}{\rho_\mathrm{b}}
      = f^\mathrm{\rho} \frac{\mathrm{d}c}{\mathrm{d} t} 
    \qquad \mathrm{with} \qquad
    f^\mathrm{\rho} = \frac{- \rho^\mathrm{\Delta}}{\rho} \frac{\partial \, m}{\partial \, c} \, . \label{equ:primal_mass_conservation_final}
\end{align}
Here $\dot\rho_\mathrm{a} = -\dot\rho_\mathrm{b}$ represents the mass transfer rates into phases $a$ and $b$. Mass conservative CH formulations yield non-solenoidal velocity fields unless $f^\rho$ vanishes \citep{kuhl2021cahn}. This, in turn, suggests employing the hyperbolic EoS $m^\mathrm{(2)}$ or its linearized variant $m^\mathrm{(3)}$, which can compress the non-solenoidal regime to a fairly small layer controlled by $\gamma^\mathrm{m}$.

\subsubsection{Governing Equations}
\label{sec:ContinousDes}
The governing equations primarily refer to the momentum and continuity equation for the mixture as well as a transport equation for the volume concentration of the foreground phase. They provide the pressure $p$, velocity $u_i$ and volume concentration $c$, viz.
\begin{alignat}{3}
    &\mathrm{R}^\mathrm{p}&&=  \frac{\partial u_k}{\partial x_k}  - \frac{f^\mathrm{\rho}}{1 + f^\mathrm{\rho}  c} \;  \frac{\partial}{\partial  x_k} \bigg[ M \frac{\partial  \psi}{\partial  x_k}\bigg] &&= 0\, , \label{equ:primal_rans_masse_unlucky} \\
    &\mathrm{R}^\mathrm{c}&&= 
    \frac{\partial c}{\partial t} + u_k \frac{\partial c}{\partial x_k} 
    - \frac{1}{1 + f^\mathrm{\rho}  c}   \; \frac{\partial}{\partial  x_\mathrm{k}} \bigg[ M \frac{\partial  \psi}{\partial  x_\mathrm{k}}\bigg] &&= 0\, , \label{equ:primal_rans_concee_unlucky} \\
    &\mathrm{R}_i^\mathrm{u_i}&&= \rho \left[ \frac{\partial u_i}{\partial t} + u_k \frac{\partial u_i}{\partial x_k} \right]
    + \frac{\partial }{\partial  x_k} \bigg[p^\mathrm{ eff}   \delta_{ik} - 2  \mu^\mathrm{eff}  S_{ik} \bigg] - \rho  g_i - f_i^{\mathrm{ST, CH}} + \nonumber \\ 
    &\qquad &&\qquad\frac{2}{3} \frac{\partial}{\partial x_i} \bigg[ \mu \frac{f^\mathrm{\rho}}{1 + f^\mathrm{\rho}  c} \; \frac{\partial}{\partial x_k} \bigg[ M \frac{\partial  \psi}{\partial  x_k}\bigg] \bigg] &&=0 \, . 
    \label{equ:primal_rans_mome_unlucky} 
\end{alignat}
The unit coordinates and the strain rate tensor are denoted by the Kronecker Delta $\delta_{ik}$ and $S_{ik} = 0.5 (\partial u_i / \partial x_k + \partial u_k / \partial x_i)$. The framework supports turbulent flows, where $u_i$ corresponds to averaged or filtered velocities and $p^\mathrm{eff} = p + p^\mathrm{t}$ is additionally augmented by a turbulent kinetic energy (TKE, $k$) term, i.e., $p^\mathrm{t} = 2 \rho k/3$. Along with the Boussinesq hypothesis, the dynamic viscosity $\mu^\mathrm{eff} = \mu + \mu^\mathrm{t}$ of turbulent flows consists of a molecular and a turbulent contribution ($\mu^\mathrm{t}$), and the system is closed by a two-equation turbulence model to determine $\mu^\mathrm{t}$ and $k$ in this study. Details of the turbulence modeling practice are outlined in Sec.\ \ref{sec:turb}. The term $f_i^{\mathrm{ST,CH}}$ refers to the 
surface tension force of the CH approach, cf.\ (\ref{equ:STCH}) and details in \citet{kuhl2021cahn}, which is replaced by a model in the present study.

Contributions arising from the two-phase CH model refer to  the respective last terms of Eqns.\ (\ref{equ:primal_rans_masse_unlucky})-(\ref{equ:primal_rans_mome_unlucky}) and the surface tension force in (\ref{equ:primal_rans_mome_unlucky}). The Partial Differential Equations (PDE) system agrees with the classical VoF framework for a vanishing mobility $M \to 0$. Moreover, an appreciated divergence-free velocity field (\ref{equ:primal_rans_masse_unlucky}) is obtained for $M\ne 0$ in combination with $f^\rho=0$. Using the nonlinear material model  $m^\mathrm{(3)}$ in (\ref{equ:eos_hyperbol}), $f^\mathrm{\rho}$ approximately vanishes due to $\partial m / \partial c \to 0$ virtually everywhere  for sufficiently small values of $\gamma^\mathrm{m}$. While this yields the neglect of net diffusion fluxes in (\ref{equ:primal_rans_masse_unlucky}),  
it leaves a diffusive term within the concentration Equation (\ref{equ:primal_rans_concee_unlucky}). The latter arises from the first part of the gradient of chemical potential $\psi$ in Eqn.\  (\ref{equ:CHceqB}). Such under-resolved CH-VoF methods consistently employ $f^\mathrm{\rho} \to 0 $ to simplify the PDE system, viz.

\begin{alignat}{3}
    &\mathrm{R}^\mathrm{p} &&= \frac{\partial u_k}{\partial x_k} &&= 0 \, , \label{equ:conti} \\
    &\mathrm{R}^\mathrm{c} &&= \frac{\partial c}{\partial t} + \frac{\partial u_k c}{\partial x_k} - \frac{\partial}{\partial x_k} \left[ \nu_\mathrm{c} \frac{\partial c}{\partial x_k} \right] &&= 0 \, , \label{equ:conc} \\
    &\mathrm{R}_{i}^\mathrm{ u_i} &&=  \frac{\partial \rho u_i}{\partial t} + \frac{\partial v_k \rho u_i}{\partial x_k} + \frac{\partial }{\partial x_k} \bigg[p^\mathrm{ eff} \delta_{ik} - 2 \mu^\mathrm{ eff} S_{ik} \bigg] - \rho g_i - f_i^{\mathrm{ST,BR}} &&=0 \label{equ:mome} \, .
\end{alignat}
Surface tension effects are now considered by the continuum model proposed by \citet{brackbill1992continuum}, cf.\  Sec.\ \ref{subsubsec:surface_tension}. This is motivated when decomposing the surface tension force into an isotropic (pressure-like) term and a term proportional to $C_2$, viz. 
\begin{alignat}{2}
    \label{equ:STCH}
    f_i^{\mathrm{ST,CH}} = \psi \frac{\partial c} {\partial x_i}&= C_1 \frac{\partial b}{\partial x_i} + \frac{C_2}{2} \frac{\partial}{\partial x_i} \left(  \frac{\partial c}{\partial x_k}\right)^2 &&- \frac{\partial}{\partial x_k}\left( C_2 \frac{\partial c}{\partial x_k} \frac{\partial c}{\partial x_i}\right) \nonumber\\
    &= \frac{\partial }{\partial x_i} \; p^{\mathrm{ST,CH}}  &&- \frac{\partial}{\partial x_k}\left( C_2 \frac{\partial c}{\partial x_k} \frac{\partial c}{\partial x_i}\right)\,.
\end{alignat}
The $p^{\mathrm{ST,CH}}$-term scrambles with the pressure and the second term is likely to vanish in under-resolved conditions due to $C_2 \sim \gamma_c$. The nonlinear apparent viscosity reads $\nu_\mathrm{ c} = M \partial^2 b/\partial c^2 = 2 C_\mathrm{1} \, M (6 \, c^2 - 6 \, c + 1)$, cf.\ Sec.\ \ref{sec:conctrans}. It follows from a double-well potential $b = (c-1)^2 c^2$ to be minimized in a phase separation process and locally acts either diffusive ($\nu_\mathrm{c} \ge 0$) or compressive ($\nu_\mathrm{c} <0$).
The two-phase flow model is closed by assigning the product $C_\mathrm{1} M$ to a spatially constant value that is guided by the numerical diffusion of the approximation for the convective concentration transport as suggested by \citet{kuhl2021cahn}. 

\subsubsection{Surface Tension for Under-Resolved Flows}
\label{subsubsec:surface_tension}
To consider surface tension effects in simulations that do not sufficiently resolve the thin interface, a model proposed by \citet{brackbill1992continuum} is used. The  specific surface tension force reads 
\begin{align}
    \label{equ:brackbill}
    f_i^\mathrm{ST,BR} = \sigma \kappa_\sigma \frac{\partial c}{\partial x_i} \, , 
\end{align}
where $\kappa_\sigma$ denotes the curvature of the free surface, and the surface tension is assigned to the conventional value for air-water interfaces, 
i.e., $\sigma = 0.07\,[\si{N \per m}]$ (at $T = 20\,\si{\celsius}$). The accurate calculation of the interface curvature is challenging for an interface-capturing approach. Therefore, the present study employs a suggestion of \citet{ubbink1997}, viz. 
\begin{equation}
    \kappa_\sigma = - \frac{1}{V} \oint_{S(V)}
    \frac{\left(\frac{\partial \tilde{c}}{\partial x_i} \right)}{\left| \frac{\partial \tilde{c}}{\partial x_i} \right|} \, \mathrm{d} S.
\label{eq:curv}
\end{equation}
Equation (\ref{eq:curv}) is based on a smoothed concentration field $\tilde{c}$, obtained from a Laplacian filtering operation which is repeated twice
\begin{equation}
\label{equ:laplacian}
    \tilde{c} = \mathscr{L}(\mathscr{L}(c)) \qquad {\rm with} \qquad  \mathscr{L}(c) = \frac{\sum_{\mathrm{f}} c_\mathrm{f} S_\mathrm{f}}{\sum_{\mathrm{f}} S_\mathrm{f}}.
\end{equation}
Here f indicates the faces of a control volume, and the face concentration values $c_\mathrm{f}$ are obtained from a simple linear interpolation of the adjacent cell centers.

\subsection{Turbulence Modeling}
\label{sec:turb}
A hybrid filtered/averaging DES turbulence modeling approach is employed in the present study \citep{spalart2009}. Among the various DES suggestions, we utilize the Improved Delayed Detached Eddy Simulation (IDDES) model, which agrees with a frequently employed variant of \citet{gritskevich2012development} and builds upon the popular Shear Stress Transport (SST) k-$\omega$ Boussinesq viscosity approach of \citet{menter2003ten}. To this end, the governing equations  (\ref{equ:conti}) - (\ref{equ:mome}) are supplemented by two auxiliary equations to compute the required turbulent quantities, i.e., $\mu^\mathrm{t}$ and $p^\mathrm{t} (k)$, viz.
\begin{alignat}{3}
    &\mathrm{R}^\mathrm{k}  &&= \frac{\partial \rho k}{\partial t} + \frac{\partial u_k \rho k}{\partial  x_k} - \frac{\partial }{\partial  x_k} \ \left[ \left( \mu + \sigma_\mathrm{k} \mu^\mathrm{t} \right) \frac{\partial k}{\partial x_k} \right]  - P^\mathrm{k} + c_\mu \rho k^{1.5}/L_{\rm DES} &&= 0\,, \label{equ:tke} \\
    &\mathrm{R}^\omega &&= \frac{\partial \rho \omega}{\partial t} + \frac{\partial u_k \rho q}{\partial x_k} - \frac{\partial }{\partial  x_k} \ \left[ \left( \mu +  \sigma_\omega \mu^\mathrm{t} \right) \frac{\partial \omega}{\partial x_k} \right] - \alpha \frac{\omega}{k} P^\mathrm{k} + \beta \rho \omega^2 \, &&= 0\,. \label{equ:ediss}
\end{alignat}
Here $P^\mathrm{k}=\mu^\mathrm{t} S_{ik} S_{ik}$ represents the production of TKE, $\omega$ marks a specific energy dissipation rate, and $L_{\rm DES}$ refers to a dissipation-related turbulent length scale. The Boussinesq viscosity concept is utilized to close the unresolved turbulent stresses $\rho \overline{u_i' u_k'}$:
\begin{equation}
    \rho \overline{u_i' u_k'} = \rho \frac{2k}{3} \delta_{ik} - 2 \mu^\mathrm{t} S_{ik} \, , \qquad {\rm with} \qquad \mu^\mathrm{t} = \rho \frac{k}{\omega} \; \frac{1}{\max[1, F_2 \left| S_{ik} \right| / (\omega a_1)]}.
    \label{equ:BVM}
\end{equation}
The modeling coefficients $\alpha$, $\beta$ as well as $\sigma_\mathrm{k}$ and $\sigma_\omega$ are composed of an inner (1) and an outer (2) value, i.e.,
\begin{equation}
    \label{equ:blendingF1}
    \varphi = F_1 \varphi_1 + (1-F_1) \varphi_2 \, . 
\end{equation}
The terminology distinguishes between classical inner and outer wall boundary layer regimes, where $\varphi_{1[2]}$ represents a constant inner [outer] coefficient value, and $F_1$ as well as $F_2$ correspond to blending functions of the background SST $k-\omega$ model.  

The non-zonal transition from the filtered (LES) to the Reynolds averaged (RANS) approach is primarily managed by introducing the turbulent length scale $L_\mathrm{DES}$ in Eqn.\ (\ref{equ:tke}). The latter blends a grid-independent RANS definition $L_\mathrm{RANS} \sim \sqrt{k}/\omega$ -- that varies in space and times -- with a grid-dependent filter $L_\mathrm{LES} \sim \Delta$, viz. 
\begin{equation}
    L_\mathrm{DES} =\tilde{F}_\mathrm{d} (1+F_\mathrm{e}) L_\mathrm{RANS} + (1-\tilde{F}_\mathrm{d}) L_\mathrm{LES}, \qquad \textrm{with} \quad L_\mathrm{LES} = C_\mathrm{DES} \Delta_\mathrm{mod}.
    \label{eq:turbleng}
\end{equation}
The definition of the modified filter width $\Delta_\mathrm{mod}$ employs the maximal edge length $\Delta$ of a control volume  as well as the wall-normal distance $d$, viz.
\begin{equation}
    \Delta_\mathrm{mod} = \min \left[ c_\mathrm{w} \max\left[d, \Delta \right],\Delta \right], \qquad \Delta = \max\left[\Delta_{x_1}, \Delta_{x_2}, \Delta_{x_3} \right], 
    \label{eq:deltamod}
\end{equation}
with $c_\mathrm{w}$ being an empirical constant. An overview of the employed coefficients and blending functions is provided in the Appendix. In addition to Eqn.\ (\ref{eq:deltamod}) the blending functions $F_1, F_2, \tilde{F}_\mathrm{d},$ and $F_\mathrm{e}$ also make intensive use of the wall-normal distance $d$. The definition of wall-normal distance is replaced by the instantaneous distance from the free surface, which is dynamically evaluated in a HPC-capable manner, cf.\ section below. Moreover, a sensor is used to indicate the operating mode of the turbulence closure at a given location, i.e., RANS ($\Omega_\mathrm{LES} = 0$) or LES ($\Omega_\mathrm{LES} = 1$), viz.
\begin{equation}
\label{equ:les_region}
    \Omega_\mathrm{LES} = 1 - \frac{L_\mathrm{DES}-L_\mathrm{LES}}{L_\mathrm{RANS}-L_\mathrm{LES}}.
\end{equation}

\subsubsection*{Free Surface Distance}
\label{sec:DistanceCalc}
The turbulence closure requires the distance to the nearest air-sea interface. The respective distance field is computed from the FV approximation of a Poisson equation with zero Dirichlet [Neumann] conditions along the free surface $\Gamma$ [far-field $\partial \Omega$] for a property $\tilde{d}$ $[\si{m^2}]$, viz. 
\begin{equation}
    \frac{\partial^2 \tilde{d}}{\partial x_k^2} = -1 \quad \mathrm{in} \quad \Omega
    \qquad  \mathrm{with} \qquad 
    \tilde{d} = 0 \quad  \mathrm{on} \quad \Gamma 
    \qquad  \mathrm{and}  \qquad \frac{\partial \tilde{d}}{\partial x_k} = 0 \quad  \mathrm{on} \quad \partial \Omega 
    \label{equ:fs_distance} \,
\end{equation}
as suggested by \citet{tucker1998assessment}. Since the free surface $\Gamma$, and therefore the location of boundary conditions, is grid embedded in the current approach, we manipulate the discrete equation system, cf.\ Eqn.\  (\ref{equ:momentum_numerics}), near the free surface to secure zero values along the free surface, i.e.,
\begin{align}
    A^\mathrm{\tilde{d}, P} = \alpha_\mathrm{\tilde d} 
    \qquad \qquad \mathrm{and} \qquad \qquad
    \mathcal S^\mathrm{\tilde{d}, P} = \beta_\mathrm{\tilde d} \, . 
    \label{equ:manipul}
\end{align}
Here $A^\mathrm{\tilde{d}, P}$ and $\mathcal S^\mathrm{\tilde{d}, P}$ refer to the main diagonal and the right-hand side of the algebraic equation system at a free surface cell. The identification of free surface cells to be manipulated  is assigned to the concentration value $c=0.5$. The procedure tracks sign changes of $(c-0.5)$ across the faces of the control volumes and applies the manipulation (\ref{equ:manipul}) to face adjacent cells. The parameter $\alpha_\mathrm{\tilde d}$ is assigned to a large number, e.g., $\alpha_\mathrm{\tilde d} = 10^8$, whereas $\beta_\mathrm{\tilde d}$ depends on the concentration values of the face adjacent control volumes (1,2), viz.
\begin{equation}
    \beta_{\mathrm{\tilde d},j} = 2 \alpha_\mathrm{\tilde d} \frac{| c_j - 0.5|}{|c_1 - c_2|} \qquad {\rm with} \qquad j \in [1,2].
\end{equation}
In line with \citet{tucker1998assessment} and \citet{belyaev2015variational}, the finally employed free surface distance field $d$ [m] follows from
\begin{align}
    \label{equ:normalization}
    d = \sqrt{ \frac{\partial \tilde{d}}{\partial x_k} \frac{\partial \tilde{d}}{\partial x_k} + 2 \tilde{d} } - \bigg| \frac{\partial \tilde{d}}{\partial x_k} \bigg| \, .
\end{align}

\subsection{Wind-Wave Tank}
\label{sec:windWaveTankModel}
Applications included in Sec.\ \ref{sec:Application} refer to wind-wave flume experiments of \citet{buckley2016structure,buckley2017airflow,buckley2019}, which were obtained in a flume that approximately spans $42\,\si{m}$ (length $x_1$) $\times$ $1\,\si{m}$ (width $x_2$) $\times$  $1.25\,\si{m}$ (height $x_3$). An upstream blower initiated the experimental waves to the free surface. The waves subsequently develop under strong wind forcing towards the study area of the experimental data, located approximately $23\,\si{m}$ downstream of the airflow input. Experimental data are characterized by phase-averaged mean velocities in the air phase and wave parameters, representing the characteristics of the experimental wave train at its peak frequency, cf.\ Tab.\ \ref{tab:scenarios}. Measured results mainly refer to two-component Particle Image Velocimetry (PIV) data and their post-processing. 

The numerical wind-wave tank spans only a smaller section of the experimental flume to support the feasibility of the simulations.  The considered  domain is sized by the primary wavelength reported in the experiments, cf.\ Tab.\ \ref{tab:scenarios}. The horizontal ($x_1$) extent covers $11$ wavelengths $\lambda$ and extends up to $6\,\si{m}$ for the larger investigated wavelength. The vertical ($x_3$) extension of $2.3\,\lambda$ agrees with the experimental flume ($1.25\,\si{m}$) for the larger investigated wavelength, of which $1.3\,\lambda$ ($0.7\,\si{m}$) is water wetted in still-water conditions. The lateral (spanwise, $x_2$) extent refers to  $0.4\,\lambda$, i.e., $22\,\%$ of the experimental width for the larger wavelength.

\begin{figure}[ht!]
    \centering
    \includegraphics[width=0.8\textwidth]{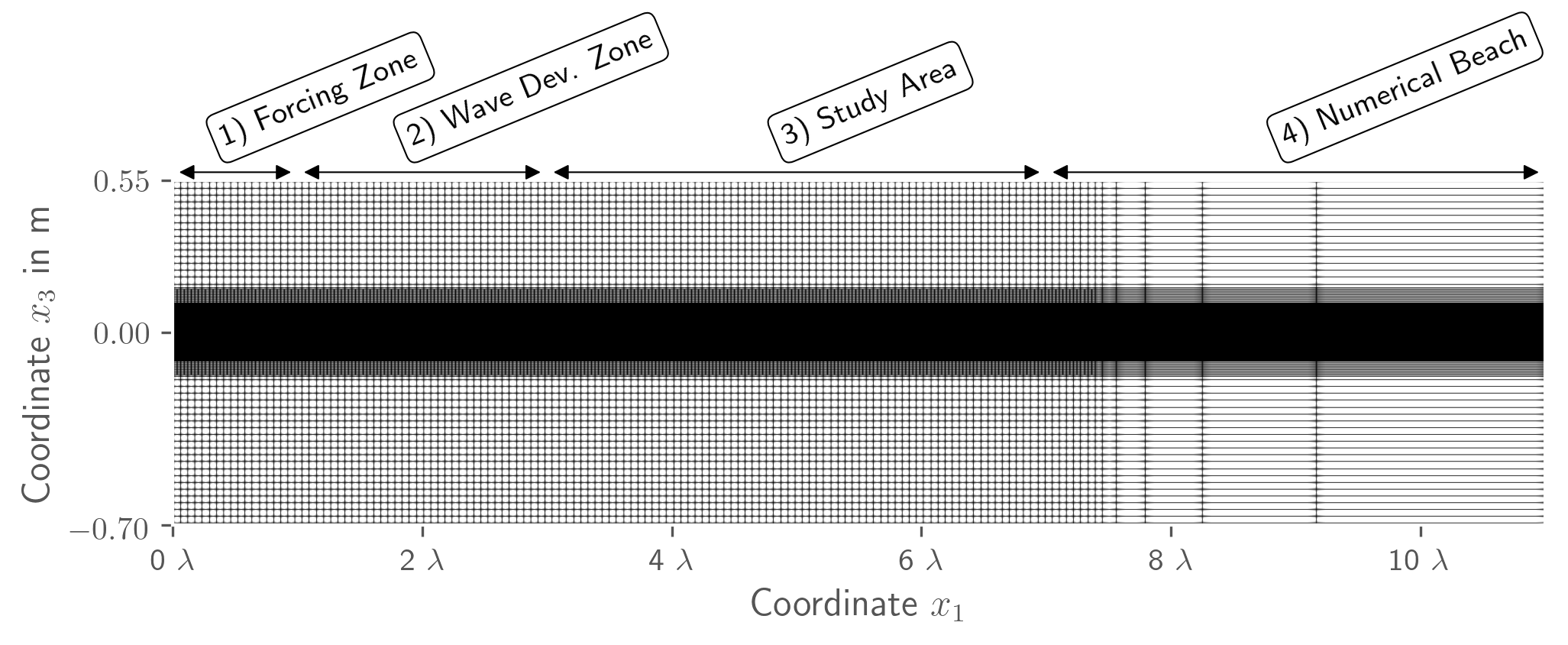}
    \caption{Illustration of the  $x_1$-$x_3$-plane of the computational domain. The calm water free surface is located at $x_3=0$ and waves travel horizontally in the positive $x_1$-direction underneath a similarly directed air stream. The domain is subdivided into a forcing zone (1), a wave development zone (2), a study area (3), and a numerical beach (4).  For an exemplary resolution of  $\Delta_{(x_1,x_2,x_3)}^\textrm{min} = [2.8, 2.8, 1.4] \cdot 10^{-3} \lambda$  the grid consists of $31.7 \cdot 10^6$ control volumes. }
\label{fig:scenario2_grid}
\end{figure}

Figure \ref{fig:scenario2_grid} illustrates the grid setup in the $x_1$-$x_3$-plane. The box-type computational domain comprises four zones, an inlet \& forcing zone ($\lambda$) at the upstream end (cf.\ Sec.\ \ref{sec:seawayBC}), followed by a wave development zone ($2\,\lambda$), a centrally located study area ($4\,\lambda$) -- where results are extracted and compared with experiments --  and a beach zone ($4\,\lambda$) to suppress wave reflections from the downstream located outlet. The grid consist of hexahedral control volumes refined around the free surface. The vertical refinement reaches a maximum resolution of $1.4 \cdot 10^{-3}\,\lambda$. The grid is substantially stretched horizontally towards the outlet in the beach zone to ensure adequate wave damping. The most crucial aspect for the comparability of experimental and numerical results is the formulation of appropriate inlet or approach flow conditions for momentum, turbulence, and surface elevation conditions (concentration) upstream of the study area described in subsequent Sections \ref{sec:windPrescription} and \ref{sec:seawayBC}. 

\subsubsection{Approaching Air-Phase Flow}
\label{sec:windPrescription}
The inlet conditions and the near-inlet forcing of the air velocities are space and time-dependent, and support considering transient, propagating orbital motion contributions to the specification of the air-phase velocity. Due to the lack of detailed scale-resolving experimental data upstream of the study area, a simple logarithmic profile based on the experimentally observed mean turbulent velocity $\bar{u}$ is adopted, cf. Sec.\ \ref{sec:tripleDecom}. In essence, detailed turbulent structures are suppressed by the considered inlet conditions, and only the dynamic mean data is imposed. The  horizontal velocity profile of the air phase $u_1^\mathrm{air}$, therefore, involves the superposition of two contributions: a smaller orbital motion contribution $u_1^\mathrm{orb,air}$, that is continuous across the air-sea interface and detailed in Sec.\ \ref{sec:seawayBC}, in addition to a horizontal logarithmic mean air-phase velocity profile $\bar{u}_1^\mathrm{air}$. Accordingly, this results in
\begin{equation}
   u_1^\mathrm{air} = \bar{u}_1^\mathrm{air} + u_1^\mathrm{orb,air} \qquad {\rm with} \qquad \bar{u}_1^\mathrm{air}= \frac{u^*}{\kappa} \ln \left( \frac{d + z_0}{z_0}\right)  \; c_\mathrm{in}\; , \qquad u_2^\mathrm{air} = 0\; , \qquad u_3^\mathrm{air}=u_3^\mathrm{orb,air} .
    \label{equ:inlet}
\end{equation}
Here $\kappa$, $u^*$, $d$, and $z_0$ refer to the Von-Karman constant, the friction velocity, the distance normal to the free surface and a roughness length. All mean-flow parameters were chosen according to the data described by \citet{buckley2019}. Moreover, a bulk-correction coefficient $c_\mathrm{in}$ is introduced to conserve a time-invariant entering air-volume flux. The air-phase velocity is dynamically imposed above the wave in a forcing zone extending one wavelength downstream from the inlet by using the implicit forcing approach described in subsequent Section \ref{sec:seawayBC}. Mind that the forcing is applied to all velocity components (even if their value is zero). 

\subsubsection{Approaching Water-Phase Flow and Implicit Forcing}
\label{sec:seawayBC}
Two water-phase-based challenges need to be addressed by the numerical wave tank, i.e., controlling the wave propagation towards the study area and suppressing upstream traveling disturbances from the outlet into the study area. Waves generation and propagation are often simulated with numerically efficient inviscid methods. More complex phenomena, such as strong wind forcing or breaking waves, require considering viscous, turbulent, and potentially two-phase flow effects. However, scale-resolving large-domain turbulent flow simulations are afflicted with prohibitive spatio-temporal resolution requirements for an accurate propagation of the incident flow towards the study area. Accordingly, small spatial domains are desirable to reduce computational effort {when periodic conditions are arguably debatable for complex flow problems. Periodic boundary conditions force wave trains to adjust to a constant wavelength, which may not correspond to the true nature of a growing wind wave and could therefore cause unfavorable disturbances, especially when the flow becomes more complex (e.g.\ breaking waves).}

Coupled viscous/inviscid methods are an attractive way out of the dilemma mentioned above. In particular, a one-way coupling between an inviscid baseline solution and the viscous solution can facilitate an efficient way for air-sea interface simulations at a moderate computational cost. In line with \citet{woeckner2010} and \citet{luo2017computation,luo2019numerical}, an appealing one-way coupling refers to an implicit solution forcing. The approach essentially manipulates the algebraic equation system to comply with a prescribed solution obtained from either nonlinear or linear wave theories for the velocity and concentration at the inlet boundaries. The manipulation of the flow is gradually relaxed towards the interior in a forcing zone that extends approximately one wavelength, cf.\ Fig.\ \ref{fig:scenario2_grid}. The present study also applies the forcing strategy to the air phase to impose the conditions described in Sec.\ \ref{sec:windPrescription}. Dirichlet conditions naturally supplement the approach along the inlet plane, cf.\ Sec.\ \ref{sec:BC}. The implicit forcing is based on an appropriate manipulation of the coefficient matrix resulting from the FV discretization of the governing Eqns.\ (\ref{equ:conc})-(\ref{equ:mome}) and proves to be very robust. The manipulation follows the suggestion outlined in Eqn.\ (\ref{equ:manipul}) and reads 
\begin{align}
\label{equ:manipulation}
 A^\mathrm{\varphi, P} \to \left(1 + \beta_\mathrm{s} \alpha_\mathrm{s} \right) A^\mathrm{\varphi, P} \, \qquad {\rm and} \qquad 
  \mathcal S^\mathrm{\varphi, P}  \to
  \mathcal S^\mathrm{\varphi, P} 
  + \left( A^\mathrm{\varphi, P} \beta_\mathrm{s} \alpha_\mathrm{s} \right) \,  {\varphi}^{*\mathrm{P}} \; ,
\end{align}
where  $\varphi$ and ${\varphi}^*$ refer to the manipulated variable and its prescribed inviscid solution at the discrete location\,P. 
The non-dimensional relative magnitude $\beta_\mathrm{s}$ controls the forcing intensity. As outlined by \citet{woeckner2010}, small values of $\beta_\mathrm{s}$ are sufficient due to forcing a larger region, i.e., $\beta_\mathrm{s}=10^{-2}$ in the present study. The scalar shape function $\alpha_\mathrm{s}$ controls the spatial extent of the forcing, and the present study employs a quadratic drop from unity to zero over one wavelength with increasing distance from the inlet boundary, cf.\ Fig.\ \ref{fig:scenario2_grid}.

Using the above-described forcing, any (known) wave-wind condition, i.e., any prescribed  ${\varphi}^*$ in Eqn.\ (\ref{equ:manipulation}), can be superposed to mimic the experimental conditions. The present study employs simple (monochromatic) 2D (plane) Airy waves, which rapidly expose non-linear features while propagating through the wave development zone and make applying non-linear wave prescriptions seem unnecessary. As an example, the following applies to the orbital velocities $ u_i^\mathrm{orb}$ of the air, and the sea phase employed herein, viz.
\begin{alignat}{5}
    u_1^\mathrm{orb,sea} &= a_\mathrm{p} \omega_\mathrm{p} \cos(\omega_\mathrm{p} t - k_\mathrm{p} x) \mathrm{e}^{-kd} \; , &&\quad u_2^\mathrm{orb,sea} &&= 0 \; , \quad &&u_3^\mathrm{orb,sea} &&= -a_\mathrm{p} \omega_\mathrm{p} \sin(\omega_\mathrm{p} t - k_\mathrm{p} x) \mathrm{e}^{-kd}, \\
    u_1^\mathrm{orb,air} &= -a_\mathrm{p} \omega_\mathrm{p} \cos(\omega_\mathrm{p} t - k_\mathrm{p} x) \mathrm{e}^{-kd} \; , &&\quad u_2^\mathrm{orb,air} &&= 0 \; , \quad &&u_3^\mathrm{orb,air} &&= -a_\mathrm{p} \omega_\mathrm{p} \sin(\omega_\mathrm{p} t - k_\mathrm{p} x) \mathrm{e}^{-kd},
    \label{equ:inlet_water}
\end{alignat}
where $d$ again refers to the free surface distance provided by Eqn.\ (\ref{equ:normalization}), $k_\mathrm{p}$ refers to the wave number, and the subscript $\mathrm p$ denotes the peak frequency. The imposed wave parameters are assigned to experimentally reported values for the wave amplitude $a_\mathrm{p}$, wavelength $\lambda_\mathrm{p}$, and its frequency $\omega_\mathrm{p} = 2 \pi f_\mathrm{p}$. Lateral velocities are again suppressed. Fig.\ \ref{fig:scenario12_spectra} compares measured and computed wave spectra upstream of the study area.

\subsubsection{Other Boundaries}
\label{sec:BC}
A second challenge refers to the behavior of the wave field downstream of the study area. Successive wave reflections from the far-field boundaries 
might restrict the exploitable part of transient simulations, particularly for short outlet distances. The suppression of such reflections requires significant numerical damping downstream of the study area. The latter is frequently achieved by the introduction of a damping zone (``numerical beach''), which is usually realized utilizing grid-stretching and/or diffusive approximations of the convective kinematics in Equation\,(\ref{equ:conc}). This aims at damping the waves before they reach the outlet boundary and can be combined with a simple hydrostatic pressure outlet in calm water conditions. 

\begin{table}[ht!]
    \caption{Summary of the applied boundary conditions}
    \begin{center}
	\begin{tabular}{ccccc}
		\hline
            & $x_1^\mathrm{min}$ & $x_1^\mathrm{max}$ & $x_2^\mathrm{min,max}$ & $x_3^\mathrm{min,max}$  \\ 
            \hline
            $u_i$ & Dirichlet & zero gradient & zero gradient & zero gradient  \\ 
            $p$ & zero gradient & hydrostatic pressure & zero gradient & zero gradient \\
            $c$ & Dirichlet & zero gradient & zero gradient & zero gradient  \\
            $k,\omega$ & Dirichlet & zero gradient & zero gradient & zero gradient  \\ 
            \hline
        \end{tabular}
    \end{center}
    \label{tab:BC}
\end{table}
Other boundary conditions can be taken from Tab.\ \ref{tab:BC}. As mentioned in Sec.\ \ref{sec:seawayBC}, the forcing approach is supplemented by Dirichlet conditions at the inlet ($x_1^\mathrm{min}$) for the velocity $u_i$ and the concentration field $c$. The pressure $p$ follows from Neumann boundary conditions.  At the outlet ($x_1^\mathrm{max}$), a simple hydrostatic pressure profile is used, with Neumann conditions for the velocity and the concentration field. At the lateral and vertical boundaries, zero Neumann conditions are applied for all quantities. 

\section{Data Acquisition \& Data Analysis}
\label{sec:data}
The data analysis and data acquisition strategy employs phase-averaged, wave-following coordinates, which supports the comparison with experimental data. The procedure is briefly described below. A more comprehensive discussion can be found in \citet{hara2015} or \citet{buckley2016structure}.

\subsection{Analytic Signal}
\label{sec:analyticSignal}
The data analysis is based on the phase information of each cell within the spatio-temporal domain. The phase information is retrieved from the analytic signal $\eta_\mathrm{a}$ of the smoothed wave elevation field $\hat{\eta}$. A Hilbert transformation $\mathscr{H}(\hat{\eta})$ is used to calculate this analytic signal by zeroing out the negative frequency components, turning the real-valued signal into a complex-valued signal  
\begin{equation}
    \eta_\mathrm{a} = \mathscr{F}^{-1} (\mathscr{F}(\hat{\eta}) 2\theta) = \hat{\eta} + i\mathscr{H}(\hat{\eta}).
    \label{eq:hilbert}
\end{equation}
In Eqn.\ (\ref{eq:hilbert}) $\mathscr{F}$ corresponds to the Fourier Transformation and $\theta$ to the Heaviside step function. The phase information is simply derived from the imaginary part of $\eta_\mathrm{a}$ , viz. 
\begin{equation}
    \phi = \mathfrak{Im}(\eta_\mathrm{a}) \, , 
    \label{eq:imphase}
\end{equation}
and is used to perform the decomposition of the flow quantities as outlined in Sec.\ \ref{sec:tripleDecom} below. An individual smoothed wave elevation field $\hat{\eta}$ is determined for each (post-processed) time instant. For this purpose, the horizontal calm-water ($x_1$-$x_2$) plane is discretized employing a regular 2D meta-grid in the study area region. The extensions of this meta-grid, its centroid positions, and edge lengths do not need to agree with the corresponding plane of the CFD grid, though this is virtually true in the present study. First, each CFD control volume in the study area is associated with a particular cell of the meta-grid. Subsequently, the vertical ($x_3$-) coordinate for all  control volume centroids of the study area that satisfy $d < 2 \Delta_{x_3}$ are registered at their associated meta-grid cell, where the factor of two is used to avoid empty meta-grid cells. The calculated free surface elevation of a meta-grid cell refers to an average of the registered vertical coordinates in case of multiple entries. Finally, the resulting elevation field $\eta$ is smoothed, using a Laplacian filtering operation -- similar to Equation (\ref{equ:laplacian}).

\subsection{Triple Decomposition}
\label{sec:tripleDecom}
The flow field is not analyzed by reference to the Cartesian coordinates ($x_1$, $x_2$, $x_3$) but as a function of the wave following coordinates $\zeta$ and $\phi$. The vertical coordinate $\zeta$ is assigned to the distance field $d$, also employed by the turbulence closure model, cf.\ Sec.\ \ref{sec:turb}. Furthermore, the phase information of each cell follows from the Hilbert transformation of the wave elevation field (\ref{eq:imphase}), described in Section \ref{sec:analyticSignal}.  Following the well-known Reynolds decomposition, the instantaneous velocity field  $u_i(\phi, \zeta)$ can be decomposed into a mean velocity $\langle u_i \rangle (\phi, \zeta)$ and a turbulent fluctuation $u_i'(\phi, \zeta)$, viz.
\begin{equation}
	u_i(\phi, \zeta) = \langle u_i \rangle (\phi, \zeta) + u_i'(\phi, \zeta).
	\label{equ:doubleDecom}
\end{equation}
When considering the interaction between air and sea, it is helpful to use a triple decomposition introduced by \citet{phillips1957generation}. The triple decomposition fractions the mean $\langle u_i \rangle$ into an ensemble mean $\bar{u}_i(\zeta)$ over all phases (between $-\pi$ and $\pi$) and a wave-coherent component $\tilde{u}_i(\phi, \zeta)$. The instantaneous velocity $u_i(\phi, \zeta)$ thereby follows from 
\begin{equation}
	u_i(\phi, \zeta)	= \bar{u}_i(\zeta) + \tilde{u}_i(\phi, \zeta) + u_i'(\phi, \zeta).
        \label{equ:tripleDecom}
\end{equation}
The triple decomposition can also be applied to other primitive variables, e.g., the pressure $p$. 

\clearpage
\section{Verification \& Validation}
\label{sec:Verification}
\begin{figure}[ht!]
    \centering
    \includegraphics[width=0.80\textwidth]{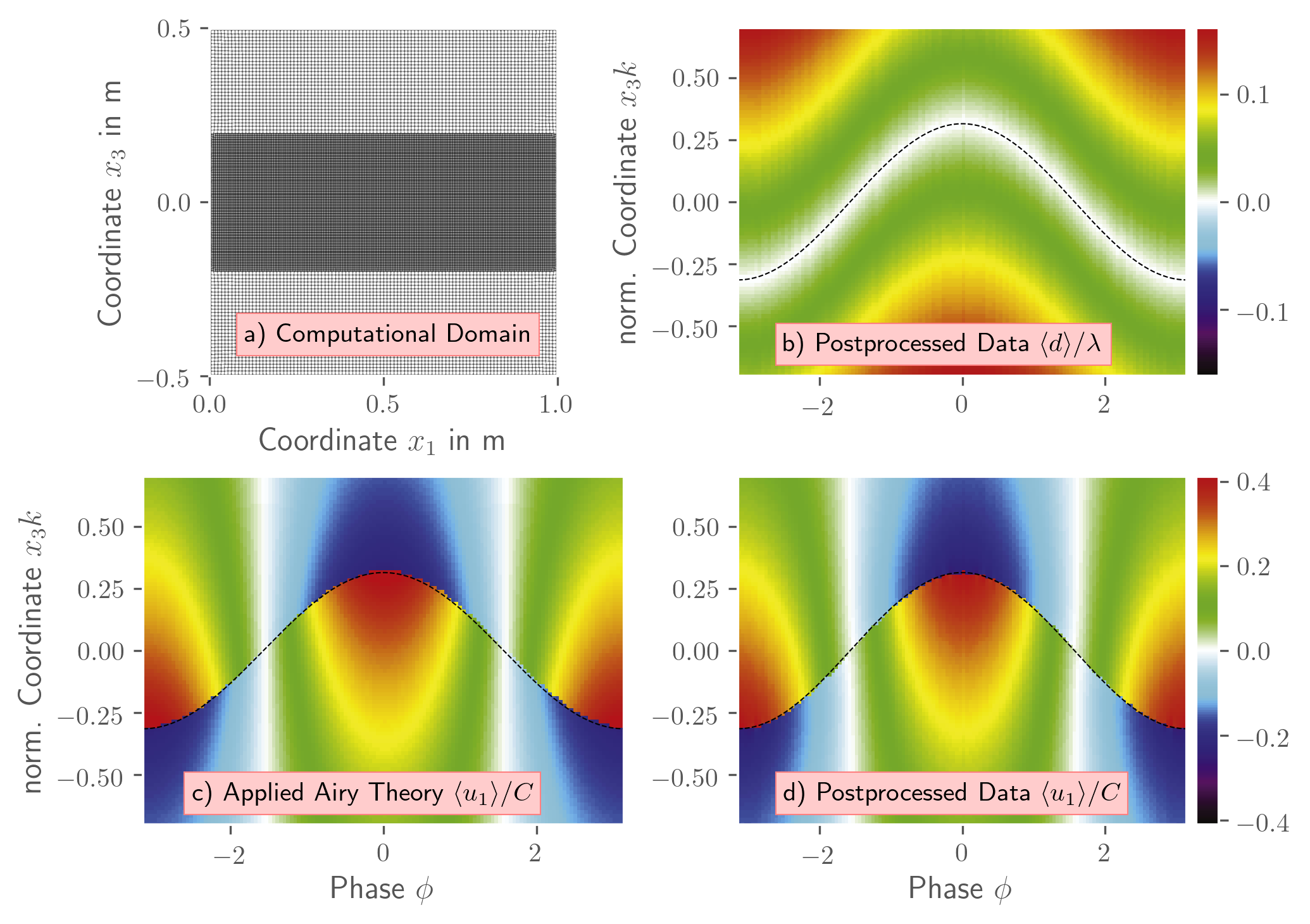}
    \caption{Results obtained for the verification case: (a) computational domain, (b) normalized post-processed distance field $\langle d \rangle / \lambda$, as well as a comparison of the prescribed (c) and processed (d) normalized velocity $\langle u_1 \rangle / C$ -- including an indication of the free surface by dashed lines.}
\label{fig:verification}
\end{figure}

The diffusive interface model and its implementation have been validated for capillary and gravity-driven two-phase flows in a previous publication by the authors \citep{kuhl2021cahn}. To verify the free surface distance calculation outlined in Sec.\ \ref{sec:turb}  and the data processing described in Sec.\ \ref{sec:data}, we analyze a simple moving wave test case in a two-dimensional domain (Figure \ref{fig:verification}). 

The velocity and concentration fields are prescribed in the entire domain by the linear wave theory using the boundary conditions described in Sec.\ \ref{sec:windPrescription} and Sec.\ \ref{sec:seawayBC}. The size of the domain reads $\lambda \times \lambda$, using a unit wavelength $\lambda = 1\,\si{m}$ and a wave amplitude of $a = 0.1\,\lambda$. Wave number $k$ and phase velocity $C$ are defined as $k = 2\pi / \lambda$ and $C = \lambda f$, with $f$ representing the frequency of the linear wave. The calm-water surface is assigned to the central height ($x_3=0$). The domain is uniformly discretized in the upper and lower regime using $\Delta_{x_1} = \Delta_{x_3} = 0.01\,\lambda$. To ensure a sufficient post-processing resolution, a refined grid with $\Delta_{x_1} = 0.005\,\lambda$,  $\Delta_{x_3} = 0.0025\,\lambda$ is employed near the free surface . 

The results of the verification test are shown in Figure \ref{fig:verification}. The processed mean velocity $\langle u_1 \rangle / C$ displays a good agreement with the prescribed data. The average distance field $\langle d \rangle / \lambda$ also offers a reasonable distribution. Minor differences result from the simplified representation of  $x_3k$ in the post-processed data, which becomes apparent by a slightly different distribution of the velocity close to the surface in both the air and the water. Instead of a more complex coordinate transformation \citep{buckley2017airflow,sullivan2000simulation}, the field data are shifted in the vertical direction only. This simplified transformation is similar to the approach used by \citet{funke2021} and \citet{husain2019boundary} and reduces the complexity of the post-processing routine. 

\clearpage
\section{Application}
\label{sec:Application}
In this study, two wind-wave scenarios (I \& II) with realistic Reynolds numbers up to $\text{Re}_\lambda \approx 20,000$ are computed. Both scenarios are based on a previous experimental investigation reported in \citet{buckley2016structure, buckley2017airflow}, where the air-sea interface is subject to strongly forced conditions. The experimentally reported results refer to airflow velocities obtained from Particle-Image-Velocimetry (PIV) measurements. They have recently been supplemented by pressure data compiled from the measured velocities of the first (I) and a slightly different third (III) scenario \citep{funke2021}. {Mind that the third scenario (III) was not directly simulated but is used for comparisons of the pressure fields in Sec.\ \ref{sec:Appmean},  because it shows similar flow phenomena to scenario II while the wave ages and slopes are of comparable magnitude.}

The experimentally investigated wind-wave spectra were narrow-banded, with clear peaks at a dominant wave frequency $f_p$. The tabulated data is used to derive the inlet conditions for the air and the sea phase as outlined in Sec.\ \ref{sec:seawayBC} and characterizes the wave fields solely with the help of the identified peak frequency $f_\mathrm{p}$. The study focuses on young waves, frequently labeled wind-waves, where the phase velocity $C_\mathrm{p}$ is of a similar order of magnitude as the phase-averaged friction velocity $u_*$ along the air-sea interface. Accordingly, this study's wave age parameter $ C_\mathrm{p}/u_*$ is relatively small, i.e., $C_\mathrm{p}/u_* < 10$, in the classification suggested by \citet{belcher1998turbulent}. Other characteristic parameters used in the study are also outlined in Tab.\ \ref{tab:scenarios}.  

\begin{table}[ht!]
    \caption{Summary of assessed wind-wave conditions for scenarios I \& II (III), where $a_\mathrm{p}$ refers to the characteristic peak amplitude, $\lambda_\mathrm{p}$ denotes the related peak wavelength, $k_\mathrm{p}$ is the peak wave number, and $f_\mathrm{p}$ refers to the peak frequency of the measured  (simulated) waves. These parameters were extracted from a somewhat broader spectrum reported by \cite{buckley2016structure}, Fig.\ 3b. The air phase is characterized by the ''10-m wind speed'' $u_{10}$, the friction velocity  $u_*$ along the air-sea interface and the related roughness length $z_0$. The wave age is characterized by $C_\mathrm{p}/u_*$ and the wave slope by $a_\mathrm{p}k_\mathrm{p}$. The Reynolds number is defined by $\text{Re}_\lambda = \rho^\mathrm{air} u_* \lambda_\mathrm{p} /\mu^\mathrm{air}$. }
    \begin{center}
	\begin{tabular}{ccccccccccc}
		\hline
            Scenario  &Descriptor&$u_{10}$ [$\si{m \per s}$] & $a_\mathrm{p}$ [$\si{cm}$] & $\lambda_\mathrm{p}$ [$\si{m}$] & $f_\mathrm{p}$ [$\si{Hz}$] & $u_*$ [$\si{m \per s}$] & $z_0$ [$\si{m}$] & $C_\mathrm{p}/u_*$ & $a_\mathrm{p}k_\mathrm{p}$ & $\text{Re}_\lambda$ \\ \hline
		I & old &$2.19$& $0.15$ ($0.18$)& $0.14$ & $3.3$& $0.07$& $3\cdot 10^{-5}$ & $6.40$ &$0.07$ ($0.08$) & $6\cdot 10^2$ \\ 
		II & young &$16.63$& $2.29$ ($2.40$) & $0.54$& $1.7$& $0.67$& $5 \cdot 10^{-4}$ & $1.37$ &$0.27$ ($0.28$) & $2\cdot 10^4$\\ \hline
		(III) & - &$9.41$& $1.20$ & $0.39$& $2.0$& $0.31$& - & $2.50$ &$0.19$ &$7\cdot 10^3$\\	\hline
	\end{tabular}
    \end{center}
    \label{tab:scenarios}
\end{table}
The material properties of the air [water] phase were set at a density $\rho$ of $1\,\si{kg \per m^3}$ [$10^3\,\si{kg \per m^3}$] and at a dynamic viscosity $\mu$ of $17\cdot10^{-6}\,\si{Pa \cdot s}$ [$10^{-3}\,\si{Pa \cdot s}$]. The CH-VoF parameter $C_1 M$ of the apparent viscosity $\nu_c$ is given by a temporal and spatially constant value mentioned in Tab.\ \ref{tab:scenariosParameters}. The latter follows precursor studies on the numerical diffusion of the convective concentration transport, as outlined by \citet{kuhl2021cahn}.
Due to the larger interface curvature radius of the first scenario (I), surface tension forces are only considered for the second scenario (II), where narrow wave crests occur. To this end, the surface tension force $\sigma$ and gravitational acceleration $g$ are set to $0.07\,\si{N \per m}$ and $9.81\,\si{m \per s^2}$, respectively.

\begin{table}[ht!]
    \caption{Employed simulation parameters, i.e., apparent viscosity $C_1 M$ of the CH-VoF model (cf.\ Sec.\ \ref{sec:conctrans}), surface tension $\sigma$, time step $\Delta_t$, data averaging time  $T_\mathrm{a}$, number of data-processing wave periods $N_\mathrm{a}$ and normalized grid spacings in the refinement zone.}
    \begin{center}
	\begin{tabular}{cccccccc}
		\hline
		Scenario & $C_1 M$ [$\si{m^2 \per s}$] & $\sigma$ [$\si{N \per m}$]& $\Delta_t$ [$\si{s}$]  & $T_\mathrm{a}$ [$\si{s}$] & $N_\mathrm{a}$ & $\Delta_{(x_1,x_2)}/\lambda_\mathrm{p}$ & $\Delta_{x_3}/a_\mathrm{p}$ \\ \hline
		I & $5\cdot 10^{-5}$ & -& $1 \cdot 10^{-4}$  & $3.5$ & $11.5$ & $2.8 \cdot 10^{-3}$ & 
		$1.4 \cdot 10^{-1}$\\ 
		II & $1 \cdot10^{-3}$& $0.07$& $1 \cdot 10^{-4}$ & $3.0$ & $5.6$ & $2.8 \cdot 10^{-3}$ & $3.3 \cdot 10^{-2}$ \\ 
		\hline
	\end{tabular}
    \end{center}
    \label{tab:scenariosParameters}
\end{table}
The investigated domain, grid set-up, and boundary conditions are described in Sec.\ \ref{sec:windWaveTankModel}. For both scenarios, the grid features a resolution of $\Delta_{(x_1,x_2,x_3)}^\mathrm{min} = [2.8, 2.8, 1.4] \cdot 10^{-3} \lambda_\mathrm{p}$ at the interface resulting in  $31.7 \cdot 10^6$ control volumes. A brief grid dependency study has demonstrated the adequacy of the resolution. Mind that the grid is strongly stretched in the beach zone, so the final cell spacing satisfies $\Delta_{x_1} > \lambda_\mathrm{p}$. Depending on the scenario, the resolution in the refinement zone leads to a wide coverage of the LES spectrum, from $y^+ \approx 2$ (close to DNS, Scenario I, air phase) up to $y^+ \geq 200$ (Scenario II, water phase). The high resolution in the first scenario is also due to the low slope, since the wave height should be resolved 
by a minimum number of cells to still accurately represent the surface. Constant time steps with $\Delta_t= 10^{-4}\,\si{s}$ are employed in both cases, which resolve the peak wave period by approximately 3000 (I) and more than 5000 (II) time steps. The complete simulations covered about $23$ (I) and $15$ (II) periods, including the initial transient phase and a subsequent data averaging time $T_\mathrm{a}$. The former is complete once the initial wave train has passed through the domain. During the subsequent $3.5\cdot 10^4$ (I) and  $3.0\cdot 10^4$ (II) time steps within  $T_\mathrm{a}$, one of fifty time steps is used for the averaging process. Accordingly, the data processing involves $N_\mathrm{a}=11.5$ (I) and $N_\mathrm{a}=5.6$ (II) wave periods. 

\begin{figure}[ht!]
    \centering
    \includegraphics[width=0.80\textwidth]{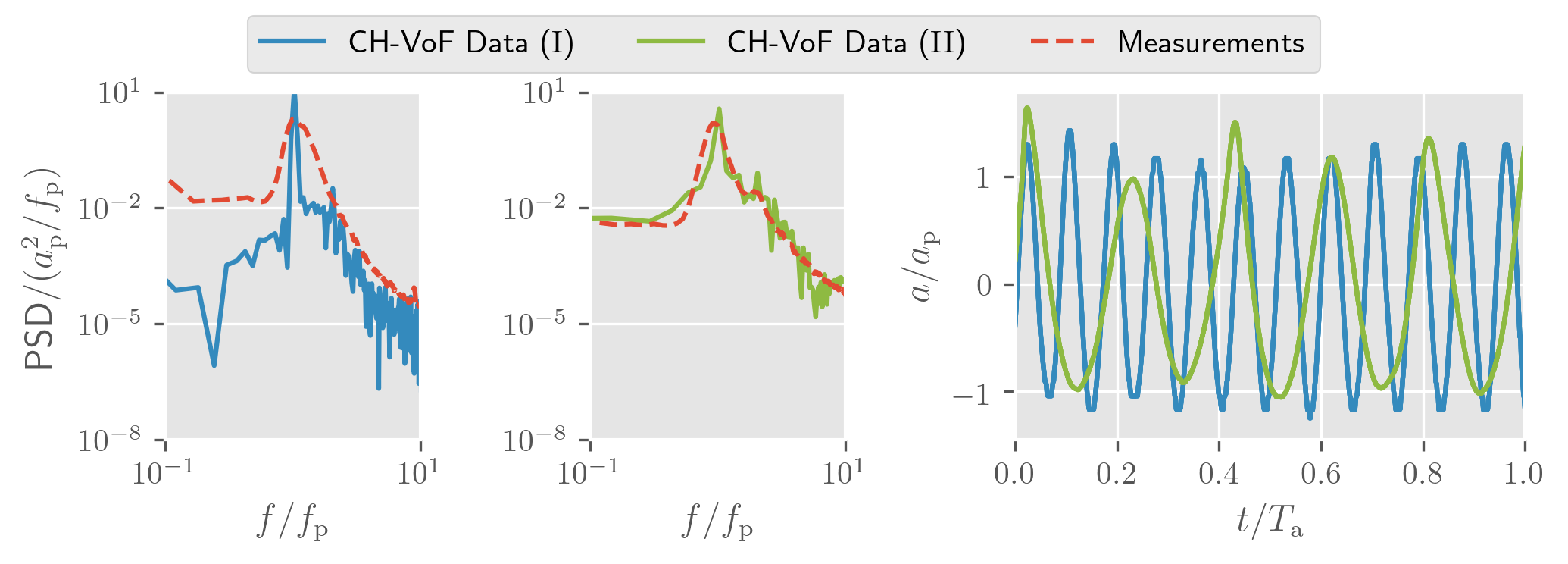}
    \caption{Normalized power spectral densities (PSD) of waves approaching the study area for the older (Scenario I; left) and younger (Scenario II; center) wave case, supplemented by the corresponding time series of the simulation data (right). Comparison of present simulations (green, blue) with measured data (red) of \cite{buckley2016structure}. All data  normalized with inlet wave parameters, cf.\ Tab. \ref{tab:scenarios}.}
\label{fig:scenario12_spectra}
\end{figure}

Exemplary normalized power spectral densities (PSD) and the corresponding time series of the waves -- extracted upstream of the study area -- are shown in Figure \ref{fig:scenario12_spectra}. PSD data are based on each time step within $T_\mathrm{a}$ and are evaluated using a basic Hann function. Compared with the measurement data in \cite{buckley2016structure}, the simulation results show a slightly narrower spectrum close the peak frequency  while the general shape is in good agreement.  However, the initially linear waves of the simulation quickly evolve into non-linear waves, with broader wave troughs and narrower crests. Note that this rough comparison of the spectra is only to ensure that the experimentally observed waves are comparable to the waves entering study area of our model. A longer evaluation period for the spectra was therefore not realized due to the computational effort.

Contour plots of instantaneous, wave coherent, and mean velocities, in addition to mean pressures and wave coherent shear stress, serve to compare numerical and experimental data in the remainder of this section. Contours are displayed in the $\langle x_1,x_3\rangle$-plane and displayed computational data is compiled from lateral averages along the homogeneous $x_2$-direction. Wave coherent quantities are calculated within the study area as a function of the wave following coordinates $\zeta$ and $\phi$, cf.\ Sec.\ \ref{sec:tripleDecom}. Moreover, we compare similarly obtained contour plots for the (resolved and modeled) computed Reynolds stresses with measured turbulent stresses and investigate the averaged horizontal velocity profile $\langle u_1(x_3) \rangle$ as well as averaged stresses along the vertical $x_3$-axis. {All data is normalized using the characteristic parameters, e.g.\ mentioned in Table \ref{tab:scenarios}. Wavelength and amplitude of the simulated waves within the study area deviate only slightly from the prescribed peak values.}

\subsection{Instantaneous Horizontal Velocity and Vorticity}
\label{sec:APPinst}
\begin{figure}[ht!]
    \centering
    \includegraphics[width=0.80\textwidth]{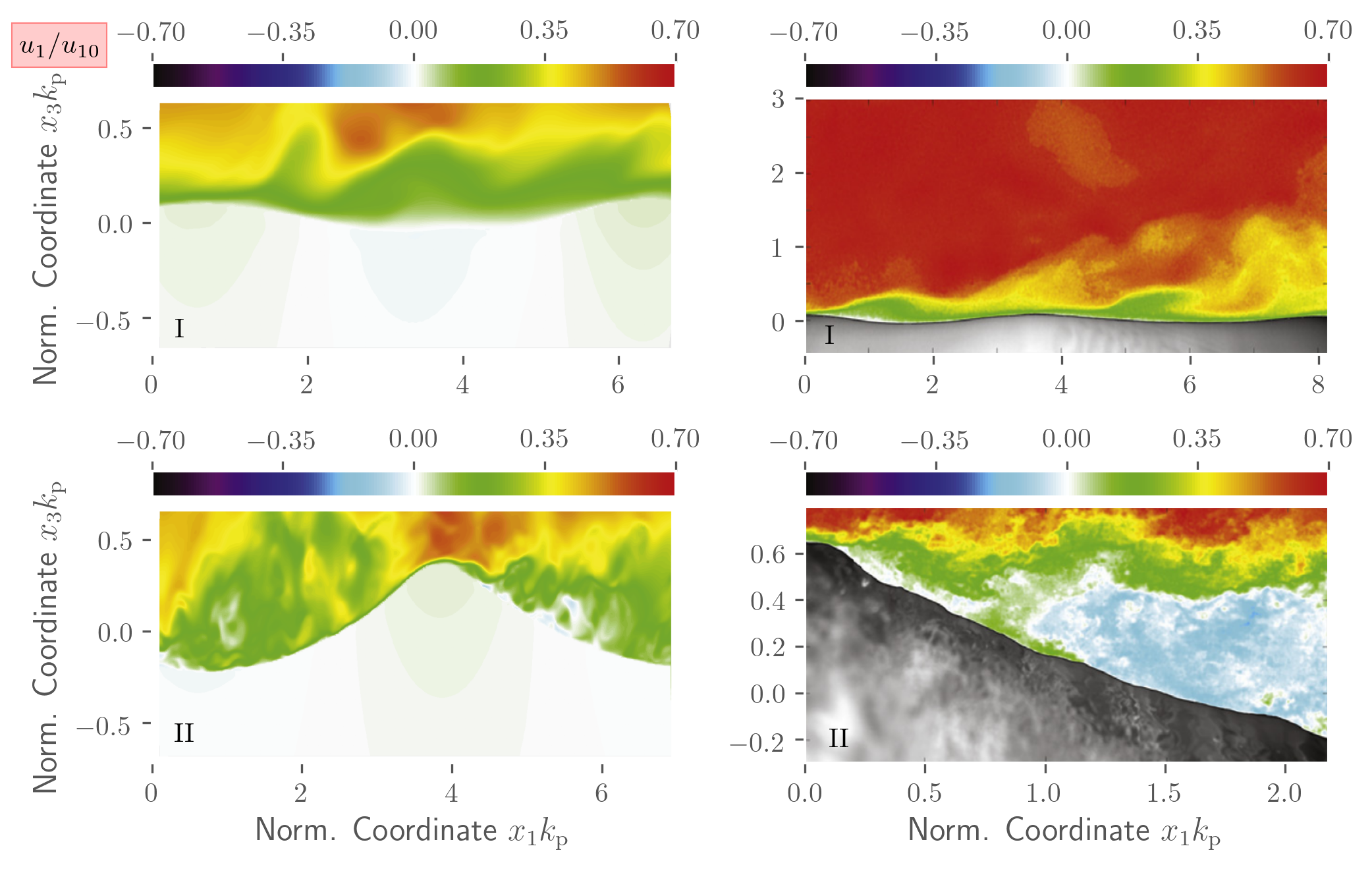}
    \caption{Comparison of (laterally-/$x_2$-averaged) instantaneous normalized horizontal velocity fields $u_1/u_{10}$ in  the study area reported from the present simulations (left) and the experiments of \cite{buckley2019} (right) for an older wave age (scenario I; top) and a younger  wave age  (scenario II; bottom).}
    \label{fig:scenario12_instantX}
\end{figure}

A first qualitative comparison of experimental and computational results is provided in Fig.\,\ref{fig:scenario12_instantX}. The figure displays snap-shot values of the predicted (left) and measured (right) normalized instantaneous  horizontal velocities $u_1/u_{\mathrm 10}$ for both wave scenarios. The top sub-figures depict the first scenario, which refers to an older wave age ($C_\mathrm{p}/u_*=6.40$), less intense wind-forcing conditions, and a smaller slope ($a_\mathrm{p}k_\mathrm{p} = 0.07$). 

\begin{figure}[ht!]
    \centering
    \includegraphics[width=0.80\textwidth]{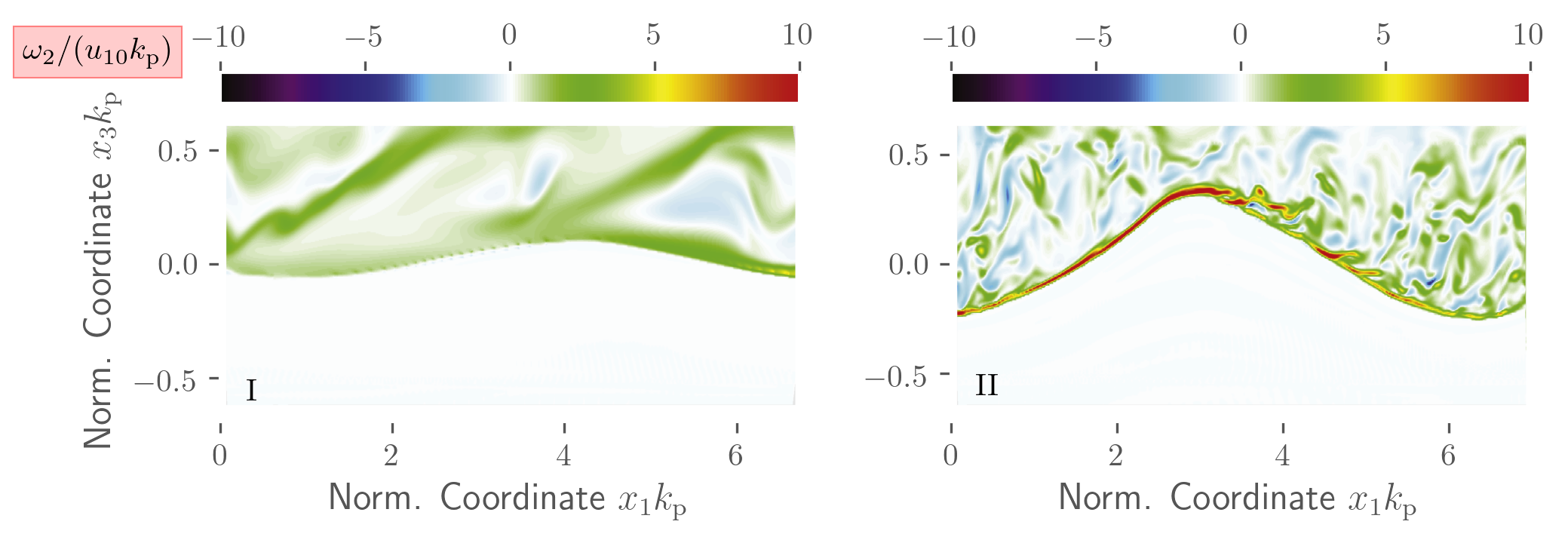}
    \caption{Comparison of instantaneous 
    normalized vorticity fields $\omega_2/(u_{10}k_\mathrm{p})$ in the study area reported from the present simulations for an older wave age (scenario I; left) and a younger  wave age  (scenario II; right).}
    \label{fig:scenario12_instantOmega}
\end{figure}

Displayed results reveal similar flow structures and  velocity values, i.e., $u_1$ increases above the crests and decreases above the troughs. In line with experiments, low-velocity fluid is ejected away from the free surface along the downwind face of the waves for the scenario I. The orbital motion is seen in the water phase prediction, which -- due to missing approach flow turbulence -- displays only very subtle  turbulence activity. As indicated in the bottom sub-figures, both simulation and experiment for the second scenario reveal significant instantaneous airflow separation behind the crest. The waves of the second, much steeper  ($a_\mathrm{p}k_\mathrm{p} = 0.27$) scenario expose substantial non-linear features. Moreover, the vorticity $\omega_2 = \partial u_1 / \partial x_3 - \partial u_3 / \partial x_1$ in the sheltered (leeward) layer of the second scenario is relatively uncorrelated, featuring islands of low vorticity, which was also reported in the experiments as an indicator for substantial airflow separation (Fig.\ \ref{fig:scenario12_instantOmega}). 

\subsection{Wave-Coherent Velocities and Shear Stress}
\label{sec:APPcohe}
To assess the influence of different wind-wave conditions on the 
mechanical energy fluxes along the air-sea interface, 
it is convenient to analyze the wave-coherent flow quantities which result from the triple decomposition introduced in Eqn.\ (\ref{equ:tripleDecom}). 

\begin{figure}[ht!]
    \centering
    \includegraphics[width=0.80\textwidth]{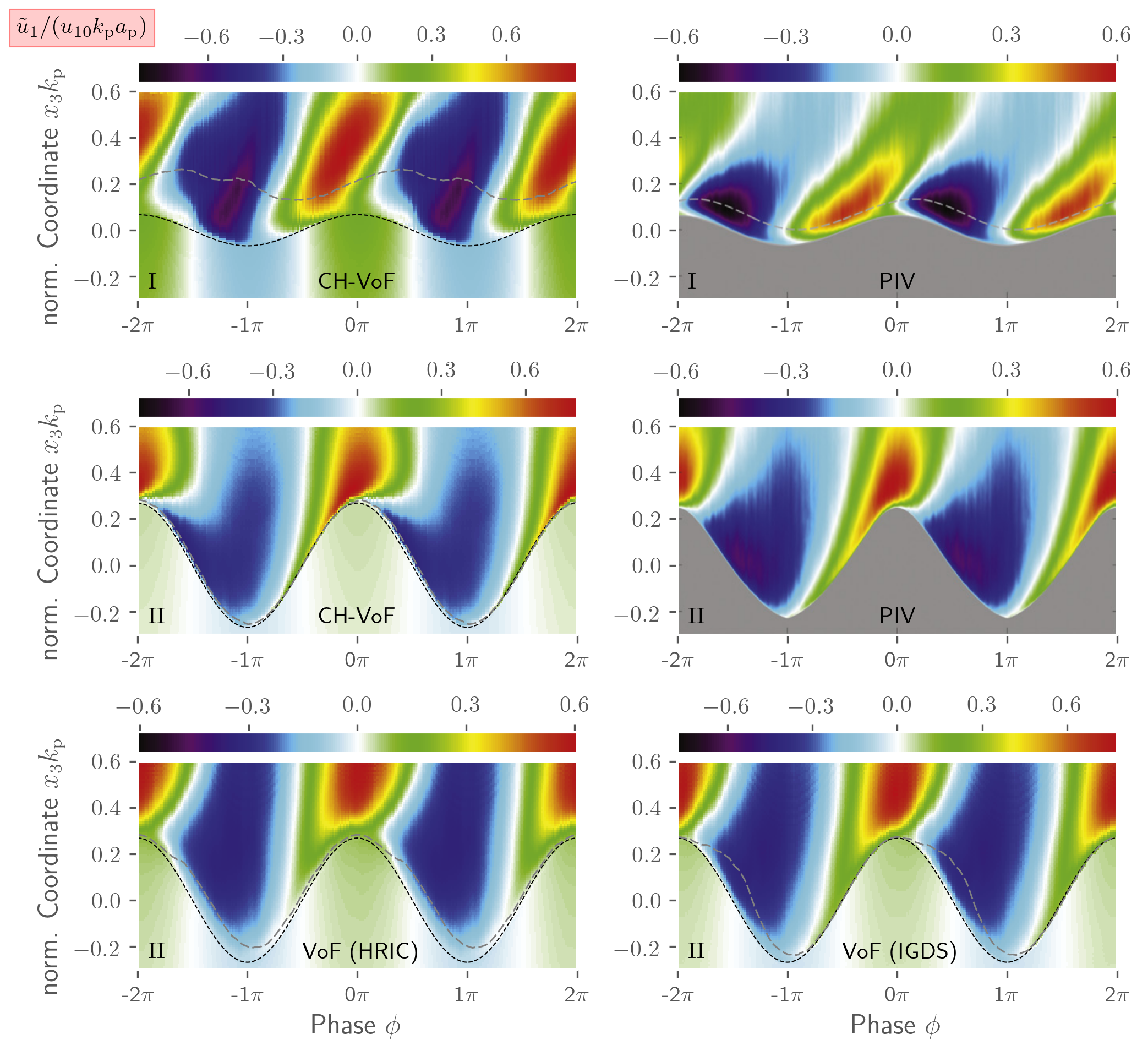}
    \caption{Comparison of (laterally-averaged) normalized horizontal wave-coherent velocity fields $\tilde{u}_1/(u_{10}k_\mathrm{p}a_\mathrm{p})$ as a function of vertical coordinate $x_3 k_\mathrm{p}$ and the phase position $\phi$. Top and center row: Results obtained from the present simulations (CH-VoF, left) and the experiments of  \cite{buckley2019} (PIV;  right) for an older wave age (scenario I; top) and a younger wave age (scenario II; center). 
    Bottom row: Supplementary results obtained from classical VoF simulations using different convective approximations: HRIC (left) and IGDS (right). Dashed grey line indicates the height of the critical layer.}
    \label{fig:scenario12_waveCoherentX}
\end{figure}

Figure \ref{fig:scenario12_waveCoherentX} depicts contour plots of the horizontal wave coherent velocity $\tilde{u}_1$ for both scenarios, which were analogously extracted from the experimental and computational data. Experimental data is  displayed in the right graphs of the top and center row, while all other sub-figures refer to computational results. Orbital motion patterns can be observed below the free surface in both scenarios. Experiments (PIV) and simulations (CH-VoF) agree that in both cases, the effect of waves is to accelerate the horizontal airflow along the upstream face of the wave and decelerate the airflow along the downstream face. For the older wave scenario (I, top), the experimental and numerical results for $\tilde{u}_1$ are in acceptable agreement. In contrast to the experimental data, the areas of extreme velocities are larger and more pronounced in the simulation results. In this regard, the agreement between numerical and experimental data improves significantly for the younger wave scenario (II, center).

\begin{figure}[ht!]
    \centering
    \includegraphics[width=0.80\textwidth]{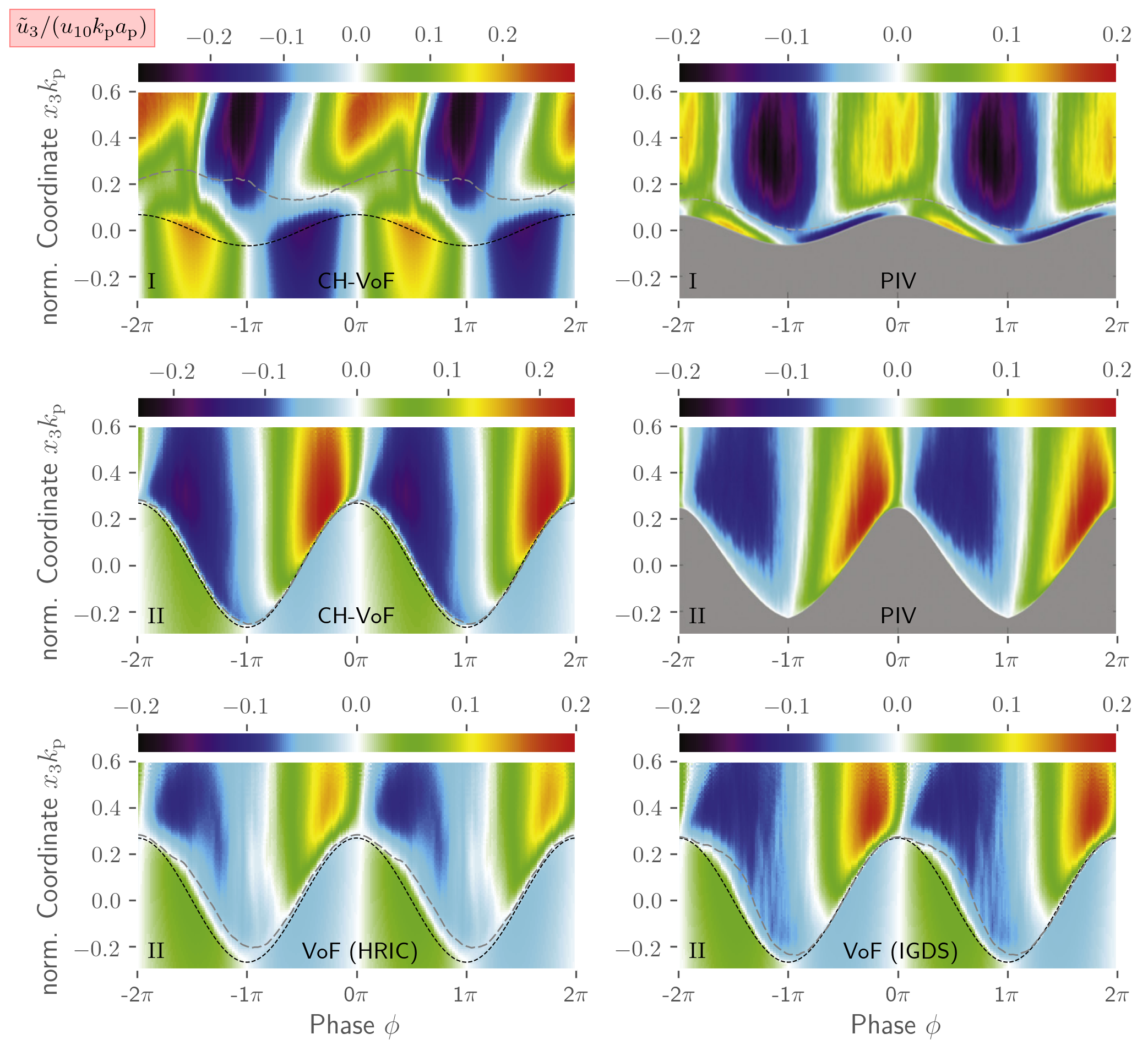}
    \caption{Comparison of (laterally-averaged) normalized horizontal wave-coherent velocity fields $\tilde{u}_3/(u_{10}k_\mathrm{p}a_\mathrm{p})$ as a function of vertical coordinate $x_3 k_\mathrm{p}$ and the phase position $\phi$. Top and center row: Results obtained from the present simulations (CH-VoF, left) and the experiments of  \cite{buckley2019} (PIV;  right) for an older wave age (scenario I; top) and a younger wave age (scenario II; center). 
    Bottom row: Supplementary results obtained from classical VoF simulations using different convective approximations: HRIC (left) and IGDS (right). Dashed grey line indicates the height of the critical layer.}
    \label{fig:scenario12_waveCoherentZ}
\end{figure}

\begin{figure}[th!]
    \centering
    \includegraphics[width=0.80\textwidth]{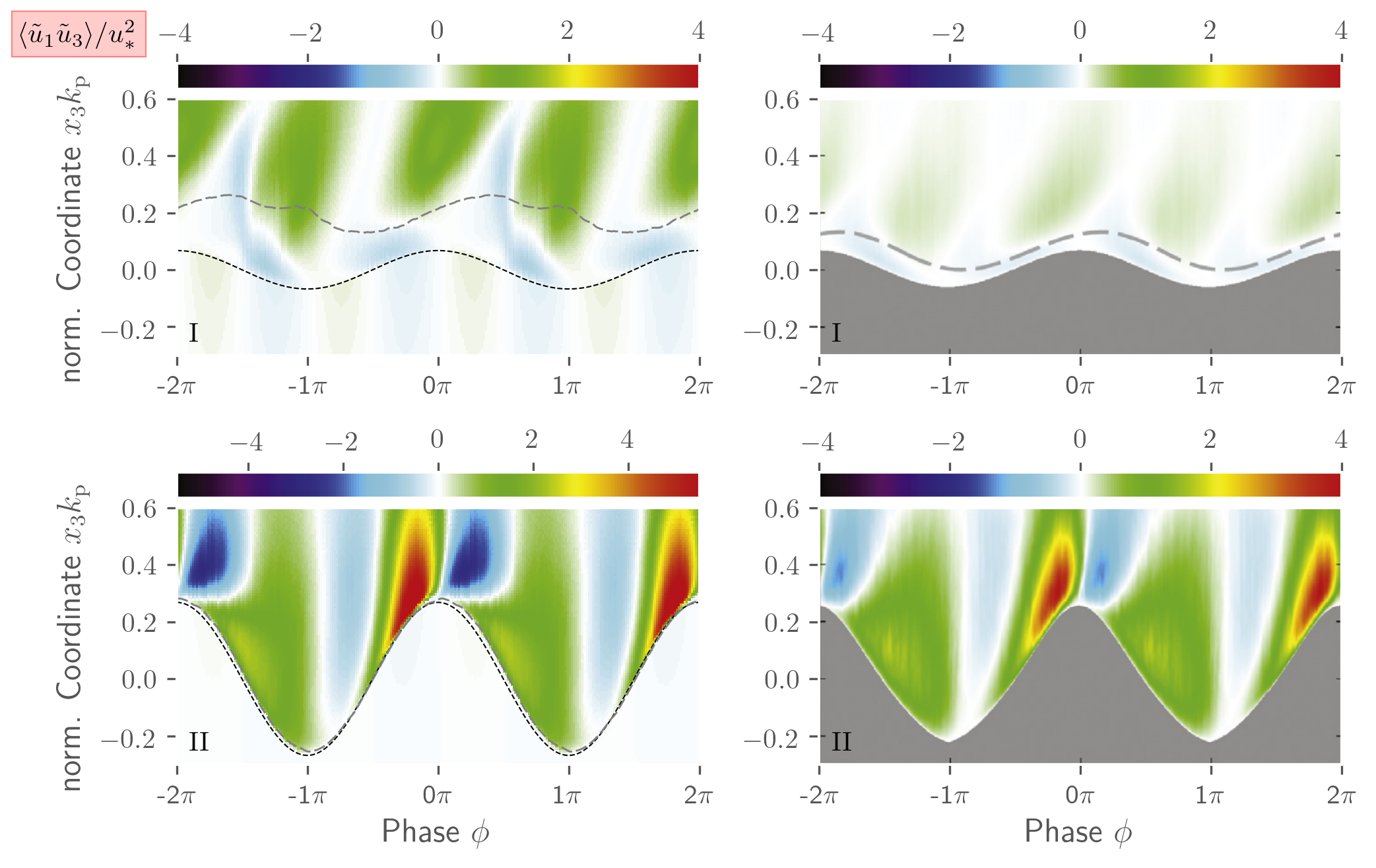}
    \caption{Comparison of (laterally-averaged) normalized wave-coherent shear stresses $\langle \tilde{u}_1 \tilde{u}_3\rangle/u_*^2$ as a function of the vertical coordinate $x_3 k_\mathrm{p}$ and the phase position $\phi$ obtained from the present simulations (left) and experiments of \cite{buckley2019} (right) for an older wave age (scenario I; top) and a younger  wave age  (scenario II; bottom). Dashed grey line indicates the height of the critical layer.}
    \label{fig:scenario12_waveCoherentXZ}
\end{figure}

For the younger wave case (II) displayed in the center of Fig.\ \ref{fig:scenario12_waveCoherentX}, the orbital motions of the air phase are hardly perceptible. On the contrary, the orbital motions clearly transit from the sea phase into the lower part of the air phase for the older wave case (I) displayed in the top sub-figures. Having entered the air phase regime in case I, the $\tilde u_1$-contours indicate a pronounced shift in the negative phase direction before turning into the positive direction above the critical layer  $x_3\ge x_3^\mathrm{crit}$. The latter is defined by $\langle u(x_3^\mathrm{crit}) \rangle  = C_{\mathrm p}$ and is indicated by the dashed grey line in each sub-figure. As confirmed by the experiments, the height of the simulated critical layer drastically increases with increasing wave age (Fig.\ \ref{fig:scenario12_waveCoherentX}). The height of the critical layer is larger above the downstream face than above the upstream face of the wave, which again outlines the accelerating/decelerating effect of the waves along the upstream/downstream face, respectively. This asymmetry reduces when increasing the wave age. For the younger wave, the critical layer is fairly thin, i.e., $x_3k_\mathrm{p}^\mathrm{max} \approx 0.02$, and thus hardly perceptible. By reference to Figs.\ \ref{fig:scenario12_waveCoherentX}-\ref{fig:scenario12_waveCoherentXZ}, it is seen that the critical layer is generally thinner in the experiment than in the simulation. This is due to a slightly different air-phase boundary layer above the wavy interface in the simulations, which is further discussed in Sec.\ \ref{sec:Appmean}. Furthermore, a positive phase shift of approximately $\pi/3$ against the wave pattern is observed for the critical layer in both simulations and experiments. 

Above the critical layer, the behavior of the velocities $\tilde{u}_1$ and $\tilde{u}_3$ changes due to the increasing influence of the applied wind forcing. 
Figure \ref{fig:scenario12_waveCoherentX} indicates that  $\tilde{u}_1$ follows an alternating pattern along the top of the study area, which is phase-aligned with the wave pattern for both investigated scenarios. When approaching the critical layer from above in Fig.\ \ref{fig:scenario12_waveCoherentX}, a phase lag is observed in both the experimental and numerical data, as indicated by the shape of the white contour lines. The maximum upstream lag of the horizontal velocity reaches approximately $\pi/3$ in the upper portion of the critical layer for the old wave case (I). Subsequently, it reduces towards zero along the interface. In line with the experimental observation, a different picture emerges for the younger wave case (II), where the upstream phase lag continuously increases towards $\pi/2$ when approaching the free surface from above, cf.\ center row of Fig.\ \ref{fig:scenario12_waveCoherentX}. For both scenarios, simulations and experiments reveal the dominance of upstream directed motion  $\tilde{u}_1$ in the trough region.

Similar observations can be made for the vertical velocity component $\tilde{u}_3$, depicted in Fig.\ \ref{fig:scenario12_waveCoherentZ}. The influence of the more pronounced critical layer is again obvious in the first scenario, where the orbital motion transits through the free surface into the air phase. When entering the critical layer from below, the contour lines of the vertical velocity bend sharply in the upstream direction between the free surface and the upper boundary of the critical layer. Above the critical layer, the situation changes, and the contour lines remain at an upstream-directed phase lag of approximately $\pi/2$ in the experiments for the older wave (I, top), which is not fully recovered by the simulations. Though the thickness of the critical layer is similarly over-predicted, the numerically determined velocity $\tilde{u}_3$ generally agrees better with the measured data for the younger wave (II, center), where magnitudes are also captured more realistically. The most distinct feature of the young [old]  wave scenario (II) [(I)] is the phase jump of $\pi$ [$\pi$/2]  observed for $\tilde{u}_3$ when crossing the free surface. Since the height of the critical layer is not negligible for the older wave case (I, top), we observe an asymmetry of the air-phase velocity above and below the critical layer height, which was also reported in the experiments and expected from Miles critical layer theory. In accordance with experiments, the computed airflow at the interface is directed down [up] for the upstream [downstream] face for the older waves (I), and an opposite trend is observed for the younger waves (II).

An interesting comparison involves the lower two rows of Figs.\ \ref{fig:scenario12_waveCoherentX} and \ref{fig:scenario12_waveCoherentZ}, which display the wave coherent horizontal and vertical velocities predicted by two classical VoF methods (bottom row) and the present CH-VoF approach (center row) for the younger wave (II). The comparison reveals  predictive improvements of the CH-VoF methods primarily related to the 
lee-ward behavior of the critical layer.
Different from the CH-VoF data, both VoF results depict 
a sudden increase of the lee-ward critical layer due to the injection of high-density fluid parts from the interface layer into the airflow. A deeper analysis of the predictive disparities between the VoF and the CH-VoF reveals a sharp but much more rippled/rugged interface of the former in the vicinity of the crest. This observation is independent of the convective approximation used for the VoF simulations (HRIC, IGDS). These small-scale ripples vary in space and time, and the flow field (occasionally) reveals strong local accelerations that are inclined against the wavy interface and promote separation. In a continuous limit, we expect similar results from a VoF and a CH-VoF framework.
Hence results of Fig.\ \ref{fig:scenario12_waveCoherentX} indicate the technical benefits of the CH-VoF approach, e.g, a minor grid dependency and a less rugged prediction of the free surface in under-resolved flows. Mind that, in contrast to the CH-VoF, the compressive face-property reconstruction schemes of many VoF methods are prone to interferences by the local Courant number and the inclination of the interface against the cell-face normal.
The maximum Courant number, however, never increased above $\mathrm{Co}=0.22$ in the present study. Moreover, the CH-VoF approach has resharpening capabilities, as indicated in \cite{kuhl2021cahn}, which can be a crucial asset of the modeling approach. Note that the EoS of the VoF simulations is in the linear form, cf.\ Eqn.\ (\ref{equ:eos_linear}). 

To assess the wave-induced momentum fluxes, we display the mean wave-coherent shear stresses $\langle \tilde{u}_1 \tilde{u}_3\rangle$ in Figure \ref{fig:scenario12_waveCoherentXZ}. Experimental and numerical results again reveal a fair predictive agreement. Simulated shear stress magnitudes are larger than the corresponding experimental values above the critical layer, particularly for the first scenario. The comparison between the two scenarios reveals a more symmetric shear stress distribution upstream and downstream of the crest near the free surface for the older wave case (I) displayed in the top sub-figures. This turns  into more obvious disparities between locations featuring similar $\phi$ 
in the critical layer and above. Following the reduction of the critical layer height, a pronounced asymmetry of the wave-coherent shear stress  around the crest is observed on the air side for the younger wave scenario (II, bottom), which is also confirmed by experimental observations. This also results in significantly different forces on the upstream and downstream faces, which drive the wave train. 
%

\subsection{Mean Horizontal Velocity and Pressure}
\label{sec:Appmean}
\begin{figure}[ht!]
    \centering
    \includegraphics[width=0.80\textwidth]{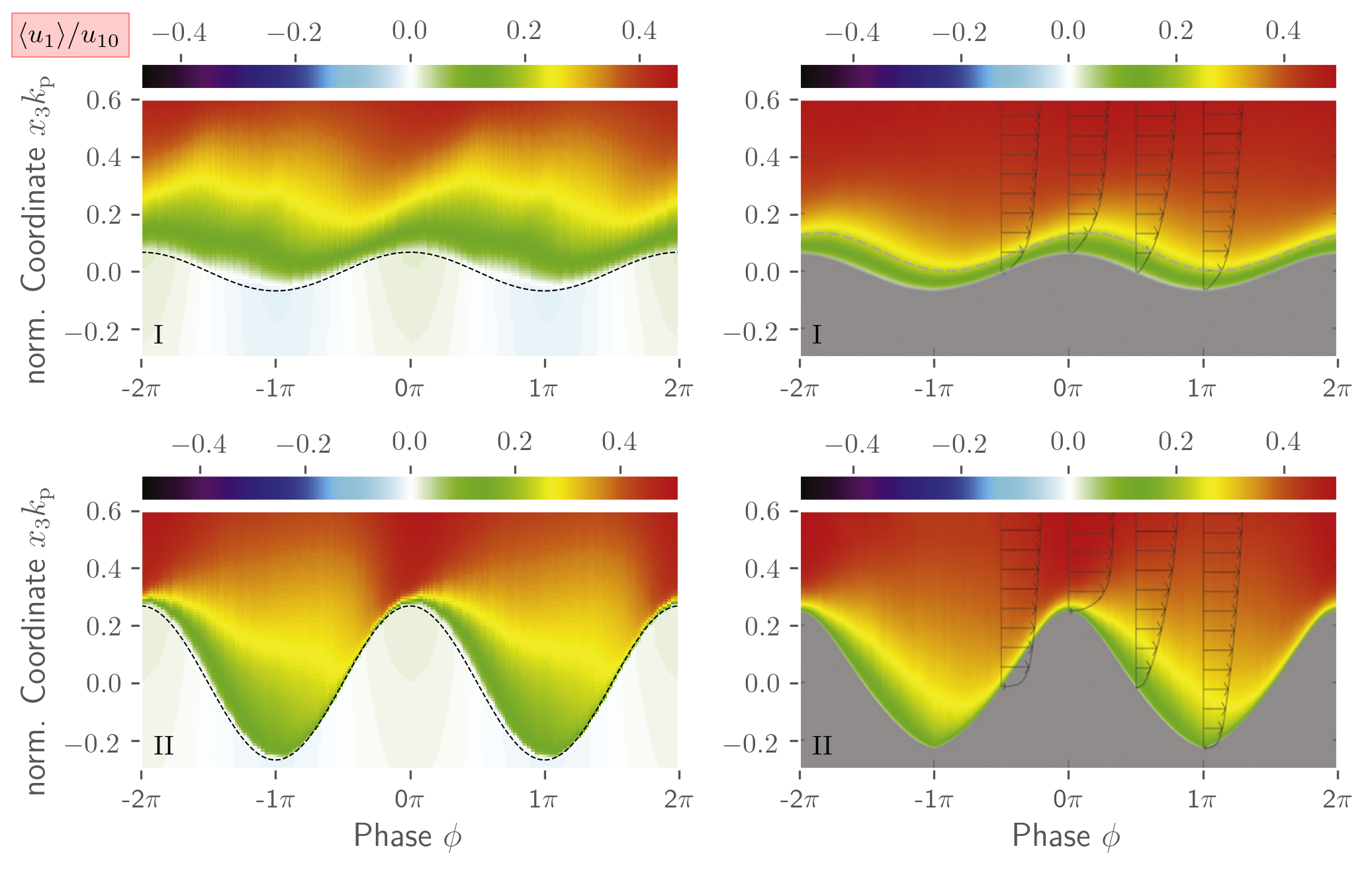}
    \caption{Comparison of (laterally-averaged) normalized mean horizontal velocity  $\langle u_1 \rangle/u_{10}$ as a function of the vertical coordinate $x_3 k_\mathrm{p}$ and the phase position $\phi$ obtained from the present simulations (left) and the experiments  of \cite{buckley2019} (right) for an older wave age (scenario I; top) and a younger  wave age  (scenario II; bottom).}
\label{fig:scenario12_meanX}
\end{figure}

Figure \ref{fig:scenario12_meanX} compares the predicted (left) and measured (right) mean horizontal velocity. While the 
far-field measurements are reproduced well, mixing-inducing (turbulent) dynamics is apparently missing in the predictions close to the free surface. 
The issue is observed in both scenarios and results in a generally thicker air-phase boundary layer above the wavy interface. However, the simulations capture the relative difference between the two scenarios, and the phase-related pattern is predicted with reasonable accuracy.

\begin{figure}[ht!]
    \centering
    \includegraphics[width=0.75\textwidth]{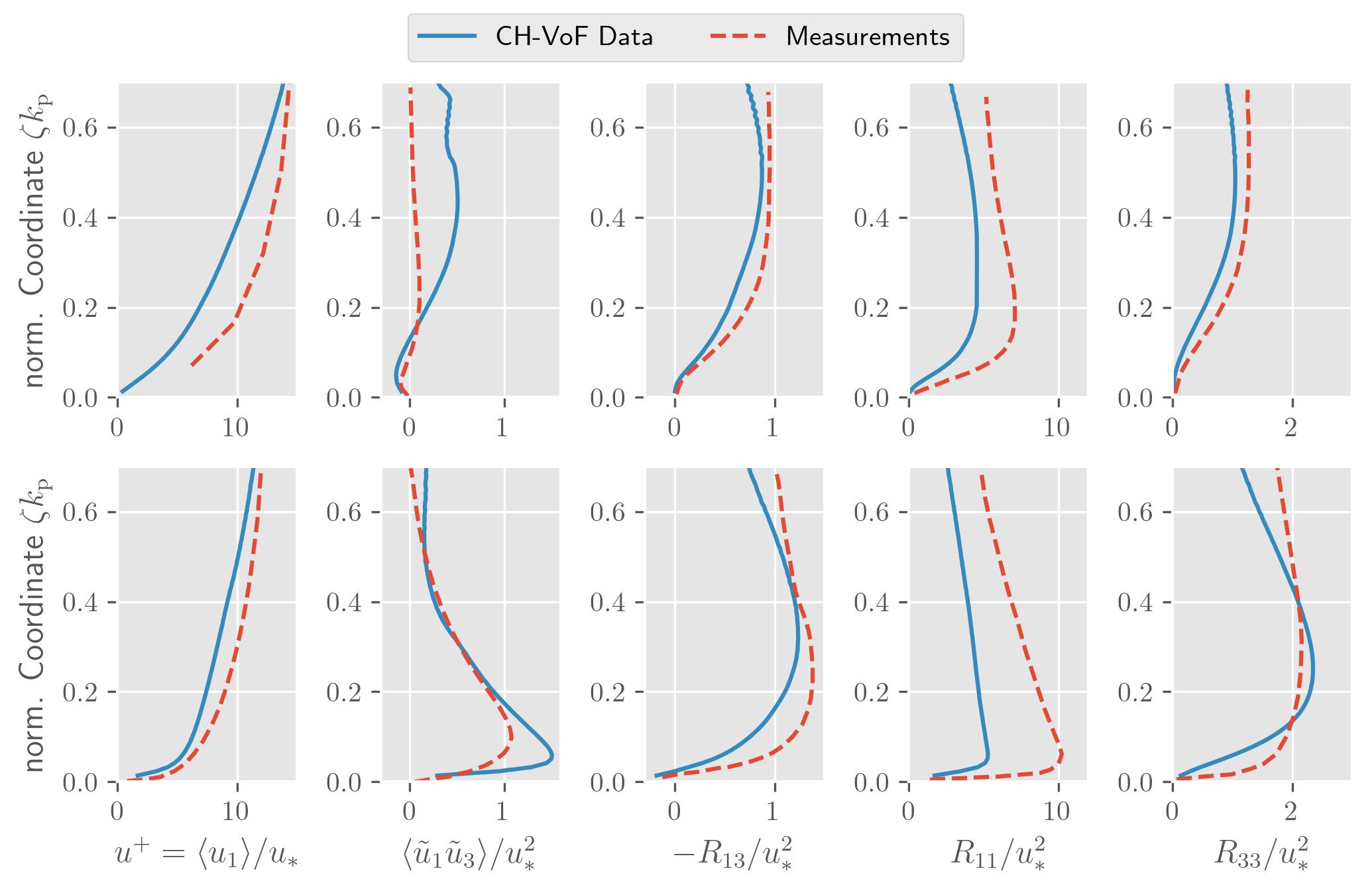}
    \caption{Comparison of predicted (blue) and measured (\cite{buckley2019}; red) vertical profiles for total mean (for all phases) quantities 
    for an older wave age (scenario I, top) and a younger wave age (scenario II, bottom); From left to right: Non-dimensional mean velocity $u^+ = \langle u_1 \rangle /u_*$, non-dimensional wave-coherent shear stress $\langle \tilde{u_1} \tilde{u_3}\rangle/u_*^2$ as well as  normalized Reynolds shear and normal stresses $-R_{13}/u_*^2$,  $R_{11}/u_*^2$ and $R_{33}/u_*^2$. }
\label{fig:scenario12_1dplots}
\end{figure}

The observation is confirmed by the above discussion on the critical layer and a comparison of the mean velocity profiles, presented on the left side of Figure \ref{fig:scenario12_1dplots}. Here, small disparities between predictions and measurements are also pronounced close to the interface for the mean velocity and the Reynolds stresses. The reason of the deficit is attributed to inconsistent and too-low levels of approach flow turbulence, which is verified by comparing the turbulent stresses in the next subsection and would suggest employing scale-resolving inlet conditions.

\begin{figure}[ht!]
    \centering
    \includegraphics[width=0.80\textwidth]{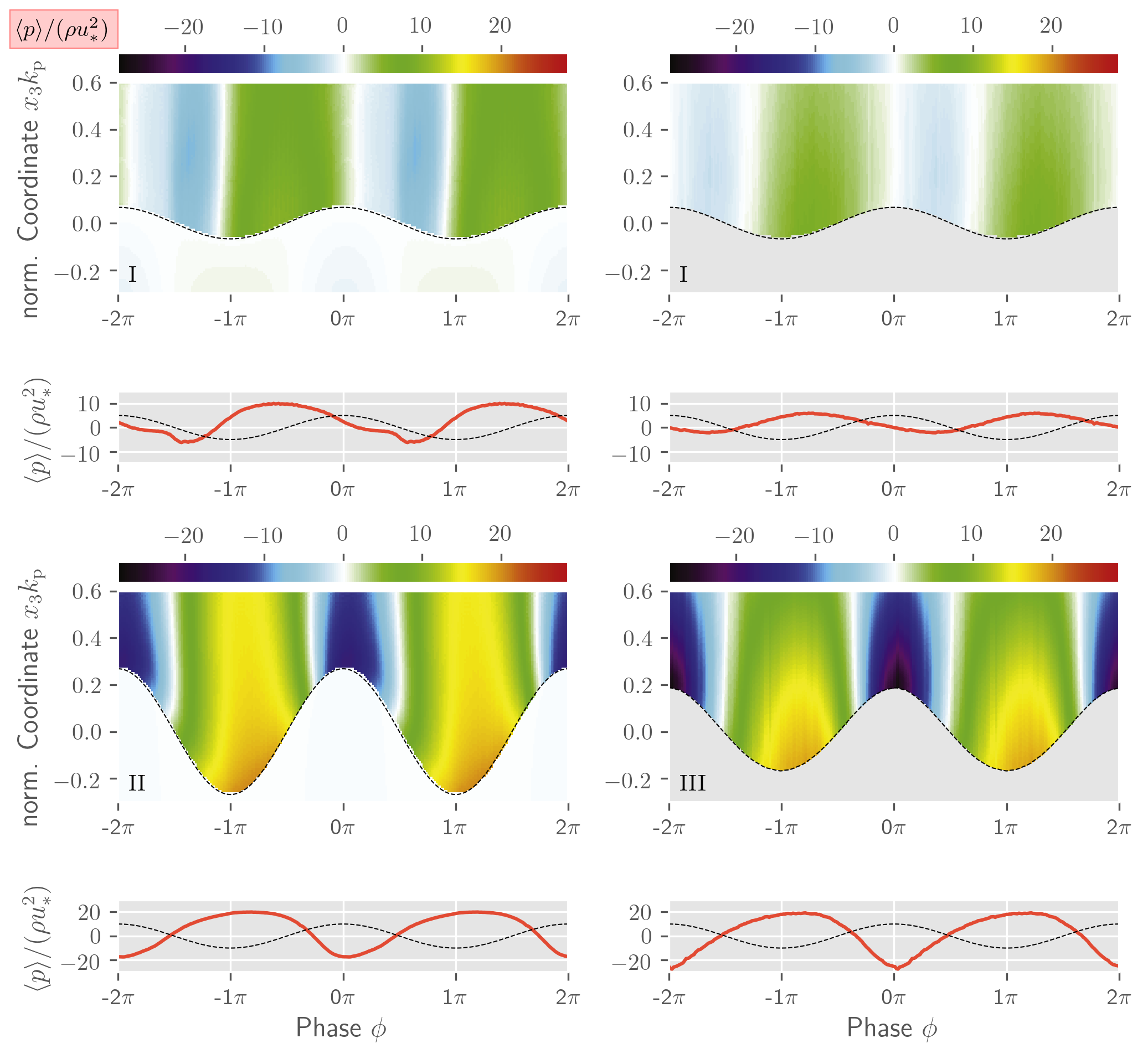}
    \caption{Comparison of (laterally-averaged) normalized averaged pressure fields $\langle p \rangle / (\rho u_*^2)$ as a function of the vertical coordinate $x_3 k_\mathrm{p}$ and the phase position $\phi$ obtained from the present simulations (left) and experiments of  \cite{funke2021} (right) for an older wave age (scenario I; top) and a younger  wave age  (scenarios II \& III; bottom). Normalized pressure distribution at the air-sea interface are shown below the contour plots.}
\label{fig:scenario12_pressure}
\end{figure}

An evaluation of the averaged pressure supplements the wave-coherent shear stress for the wave drag analysis.
The pressure contours and surface pressures displayed in Figure \ref{fig:scenario12_pressure} confirm the fair predictive agreement observed before. For the older wave age (I) shown in the top sub-figures, the upstream face of the wave is exposed to positive pressures, and the downstream face experiences negative pressure. A minor downstream phase shift of the simulated pressures is observed compared to the experimental data for this scenario, and the simulated pressure magnitudes are again slightly more elevated. Moreover, the footprint of the locally increased critical layer height near the trough for the older wave case is also visible by the small intermediate increase of the surface pressure along the downwind face. 

When attention is directed to younger wave cases (II, III) depicted by the bottom sub-figures, only a smaller regime in the vicinity of the crest is exposed to negative pressures, and the negative pressure regime is tilted slightly downstream in  the experiments and simulations. Moreover, measured and predicted surface pressures are in remarkable agreement. Due to the different wave scenarios (II, III), we omit discussing further details of the pressure contours herein. 

{The form drag per unit area  $\tau^\mathrm{D}$ is approximated by a well known relation, derived from a horizontal momentum budget  \citep{funke2021}, viz.}
\begin{equation}
    \tau^\mathrm{D}=  \left\langle p \frac{\partial \hat{\eta} }{\partial x_1} \right\rangle,
\end{equation}
with $\hat{\eta}$ representing smoothed surface elevation, cf.\ Sec.\ \ref{sec:analyticSignal}. In contrast to the averaging in \citet{funke2021}, the brackets denote to the phase averaging process as outlined in Eqn.\ (\ref{equ:doubleDecom}) and not to an average along the wave length. The computed non-dimensional form drag $\tau^\mathrm{D}/(\rho u_*^2)$ reads $0.29$ [$0.79$] for the first [second] scenario. Although the form drag is only an estimate due to the phase average, the values approximately agree with literature published data for similar wave ages and slopes discussed in \citet{wu2022} and \citet{funke2021}. 

\begin{figure}[ht!]
    \centering
    \includegraphics[width=0.80\textwidth]{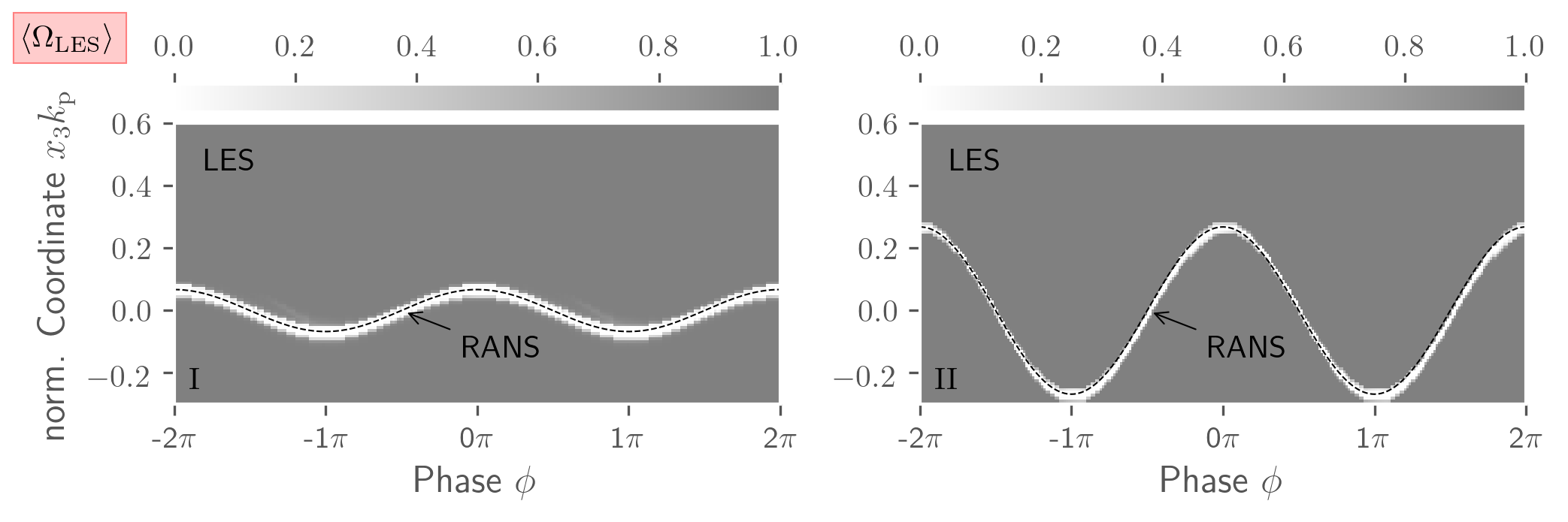}
    \caption{Comparison of 
    LES indicator function $\langle \Omega_\mathrm{LES} \rangle$ as defined in Eqn.\ (\ref{equ:les_region}) (i.e., $\langle \Omega_\mathrm{LES} \rangle= 1 \to $ LES;  $ \langle \Omega_\mathrm{LES} \rangle = 0 \to $ RANS) as a function of the vertical coordinate $x_3 k_\mathrm{p}$ and the phase position $\phi$ for an older wave age (scenario I; left) and a younger  wave age (scenario II; right).}
\label{fig:scenario12_les}
\end{figure}

Unlike all previous investigations, e.g., \cite{hara2015}, \cite{sullivan2018turbulent}, \cite{husain2019boundary} or  \cite{wu2022} the present study employs a hybrid RANS-LES (DES) turbulence modeling approach. The advantage of the method is the lower resolution requirement to compute high Reynolds number flows, all the more so when future applications of  field data measurements turn out to be more complex than an experimental wind-wave flume. As indicated by Fig.\ \ref{fig:scenario12_les}, the turbulence model is in LES mode in almost all areas of the computational domain, and the RANS mode is only activated near the interface and below where the interface is diffusive. Therefore, on average, the RANS layer only spans about three cells in the vertical direction for both scenarios.

\clearpage
\subsection{Reynolds Stresses}
\begin{figure}[ht!]
    \centering
    \includegraphics[width=0.80\textwidth]{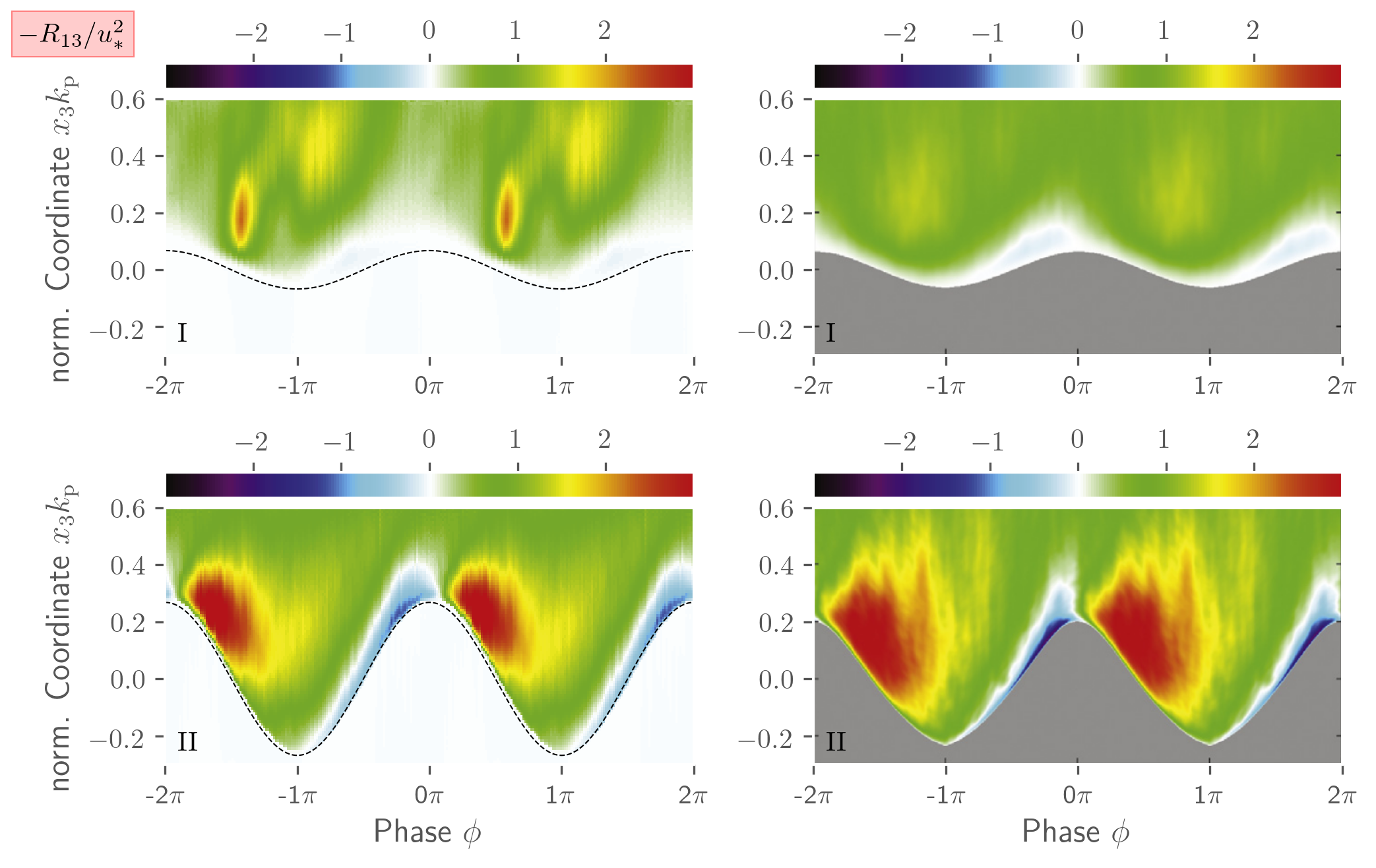}
    \caption{Comparison of (laterally-averaged) normalized Reynolds shear stresses $-R_{13}/u_*^2$ as a function of the vertical coordinate $x_3 k_\mathrm{p}$ and the phase position $\phi$ obtained from the present simulations  (left) and the experiments (\cite{buckley2019}; right) for an older wave age (scenario I; top) and a younger  wave age  (scenario II; bottom).}
\label{fig:scenario12_stressesXZ}
\end{figure}

The section compares measured and predicted Reynolds stresses $R_{ik}$. The simulated  Reynolds stresses consist of two components, the modeled stresses $R_{ij}^\mathrm{mod}$ and the resolved stresses   $R_{ij}^\mathrm{res}$, viz.
\begin{equation}
    \begin{split}
    R_{ik} &= R_{ik}^\mathrm{mod} + R_{ik}^\mathrm{res} \\
        &=2 \nu^{\rm eff} S_{ik} - \frac{2}{3} k \delta_{ik}+(\overline{u_i' u_k'})^{\rm res}.
    \end{split}
\end{equation}
Contour plots of the normalized shear stresses $-R_{13}/u_*^2$ are depicted in Figure \ref{fig:scenario12_stressesXZ}. Both cases presented herein exhibit negative (upward-directed) stresses along the upwind face of the waves. Along the downwind face of the waves, neutral (I) or positive (downward-directed, II) stresses occur, as captured by both experiments and computations.

For the first scenario, peak values of the numerical results exceed peak values reported from the experiments, but the computed stress levels show a fairly similar distribution to experimental values in the bulk of the flow. 
In contrast, the stress levels near the interface along the leeward and windward sides are slightly below the experimental stresses, confirming the observation from Fig.\,\ref{fig:scenario12_1dplots}. With attention directed to the younger waves at the bottom of Fig.\ \ref{fig:scenario12_stressesXZ}, this disparity between simulations and experiments at the interface becomes more pronounced on the windward side of the waves. However, the general features of the two shear-stress fields are (again) very similar. For the younger waves, on average, an intensification of the turbulent stresses on the leeward side of the waves is observed by the experiments and (to a slightly smaller extent) the simulations. 

\begin{figure}[ht!]
    \centering
    \includegraphics[width=0.80\textwidth]{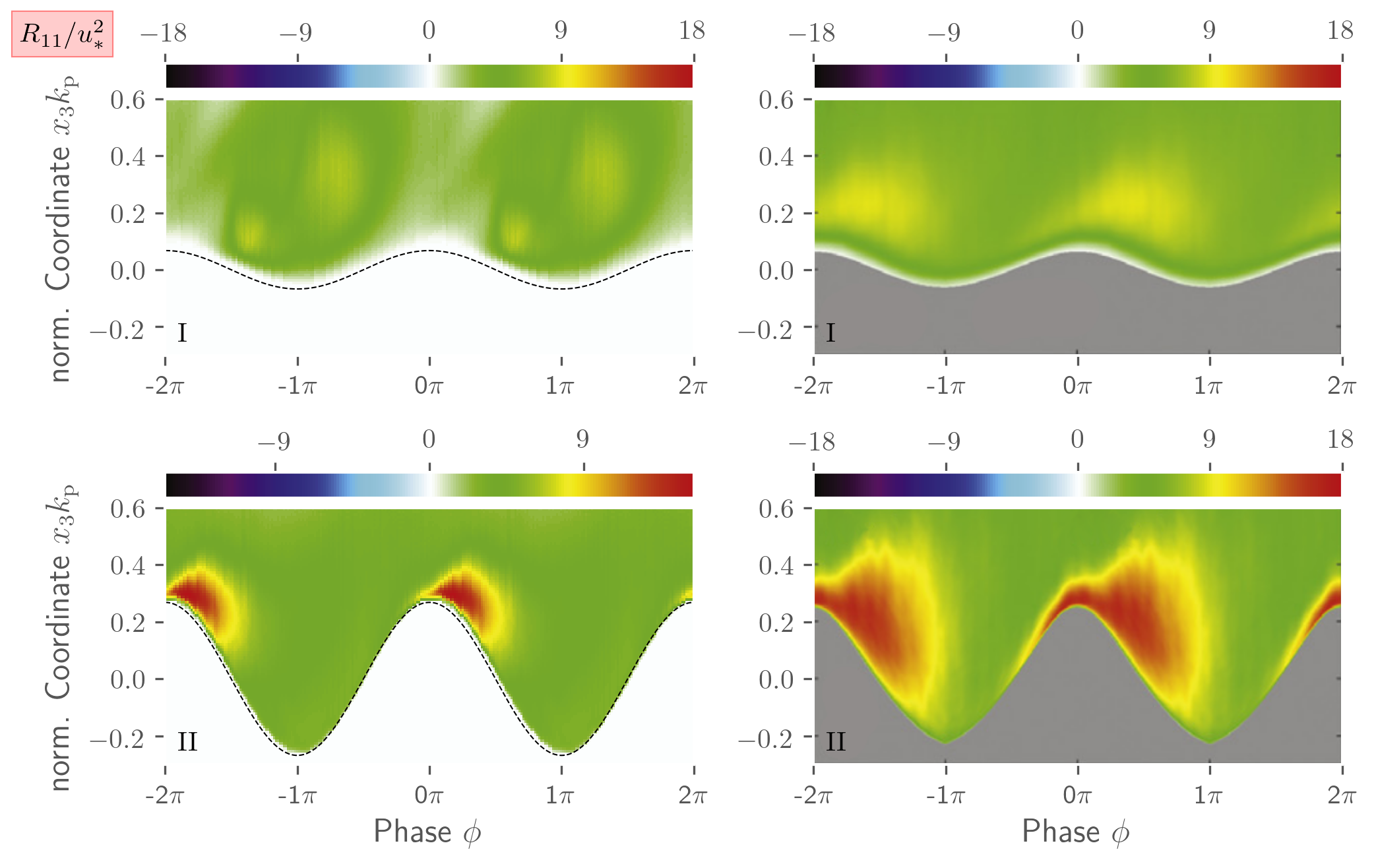}
    \caption{Comparison of (laterally-averaged) normalized Reynolds normal stresses $R_{11}/u_*^2$ as a function of the vertical coordinate $x_3 k_\mathrm{p}$ and the phase position $\phi$ obtained from the present simulations  (left) and the experiments (\cite{buckley2019}; right) for an older wave age (scenario I; top) and a younger wave age  (scenario II; bottom).}
\label{fig:scenario12_stressesXX}
\end{figure}

\begin{figure}[ht!]
    \centering
    \includegraphics[width=0.80\textwidth]{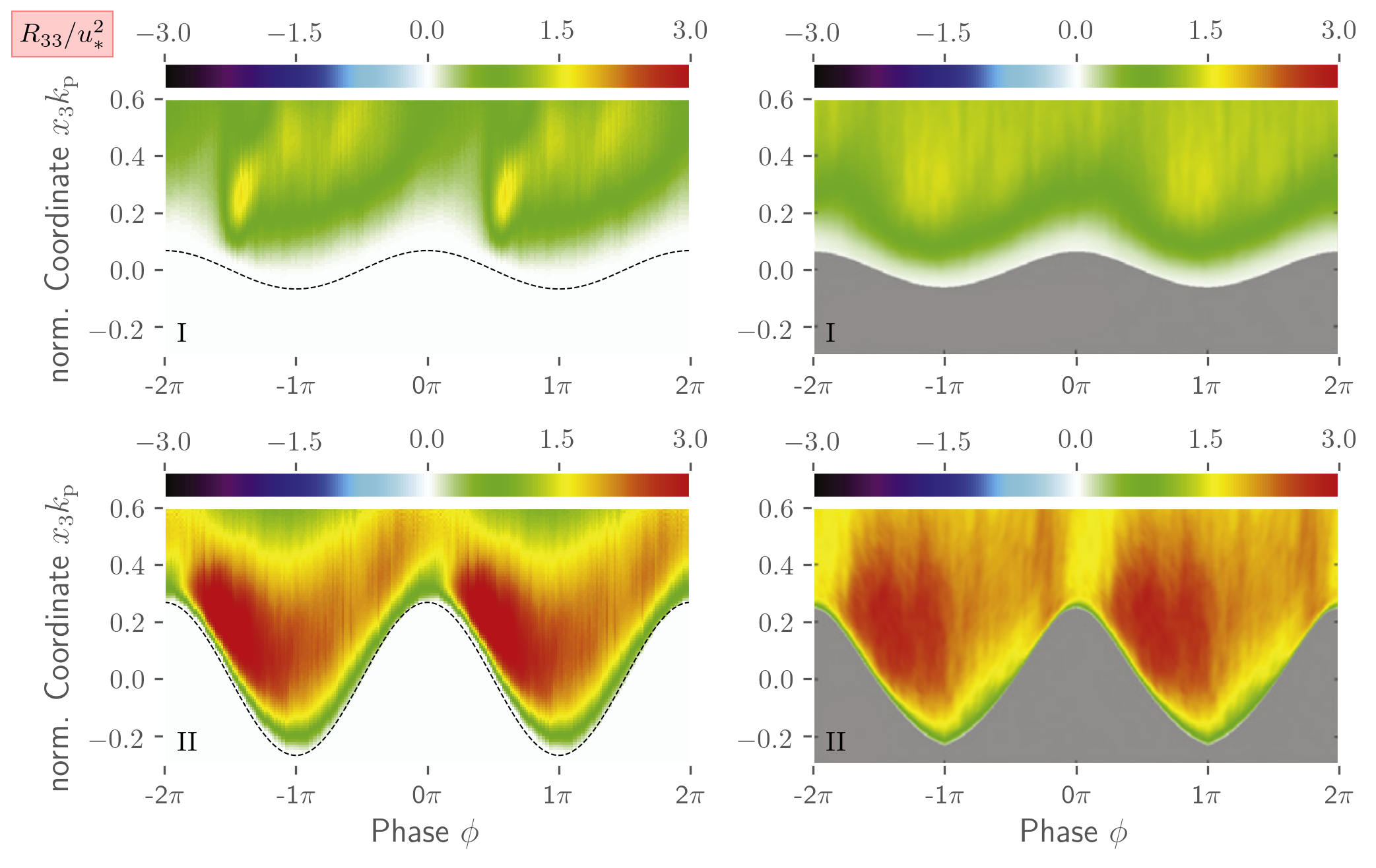}
    \caption{Comparison of (laterally-averaged) normalized Reynolds stress normal $R_{33}/u_*^2$ as a function of the vertical coordinate $x_3 k_\mathrm{p}$ and the phase position $\phi$ obtained from the present simulations  (left) and the experiments (\cite{buckley2019}; right) for an older wave age (scenario I; top) and a younger wave age  (scenario II; bottom).}
\label{fig:scenario12_stressesZZ}
\end{figure}

The missing description of the approach flow turbulence becomes more apparent when comparing the horizontal and vertical normal stresses $R_{11}/u_*^2$ and  $R_{33}/u_*^2$ in Figures \ref{fig:scenario12_stressesXX} and \ref{fig:scenario12_stressesZZ}. The predicted $R_{11}/u_*^2$ fields agree qualitatively for the older wave (I) with experiments. However, above the wave's crest on the windward side, the stress levels are underestimated. This becomes more significant for the younger wave case (II), where the color bar has been adjusted to improve the comparison. For both scenarios, the simulated stresses on the leeward side of the waves are less homogeneous than in the corresponding experimental fields. 

The vertical normal stresses $R_{33}/u_*^2$ shown in Fig.\ \ref{fig:scenario12_stressesZZ} reveal a better agreement between experimental and numerical data, though the stress level is generally much smaller than for the horizontal normal stresses. Again, differences are pronounced near the free surface, especially for the second scenario on the windward side of the waves. Mind that only minimal levels of turbulent dynamics are observed in the water phase during the simulations, which might also be a reason for the differences between the measurements and the predictions. A more detailed description of the incoming flow turbulence seems necessary to better reproduce realistic near-surface dynamics of the experiments.

\section{Conclusion \& Outlook}
\label{sec:Conclusion}
This paper scrutinizes the ability of a diffusive interface (CH-VoF) two-phase flow model to capture air-sea interface phenomena. The model comprises a pressure-based, second-order accurate FV procedure that involves a hybrid filtering/averaging (DES) approach to model flow turbulence, and is capable of simultaneously simulating atmosphere and ocean coupling through the surface wave field at realistic Reynolds numbers with surface tension. 

Attention is devoted to two different wind-wave conditions, referring to a fairly young wave age ($C_{\mathrm p}/u_*=1.37$) and a slightly older wave age ($C_{\mathrm p}/u_*=6.40$) with different wave slopes, where extensive experimental results were reported in a series of papers by \citet{buckley2016structure,buckley2017airflow,buckley2019} and \citet{funke2021}. 
Generally, the predictions are in good agreement with measurements, though the height of the critical layer is over-predicted in both cases and the shape reveals small inaccuracies above the trough for the older wave. The young (steep) wave case features a very thin critical layer and a large sheltering effect, where turbulent stresses are significantly augmented compared to the surrounding flow. The prediction of these phenomena is not as pronounced as the experimental data indicates. Nonetheless, the computations recover the (intermittent) occurrence of airflow separation downstream of the young wave crest, in line with a more disordered vorticity field, and also display a related increase of turbulence intensity downstream of the crests produced by detached shear layers. For the less steep, older wave case, the height of the predicted critical layer substantially increases, as also reported by the experiments. Similar to the experiments, the wave coherent air velocities below the critical layer start to display an asymmetry with respect to the airflow above the critical layer. As the wave gets older, we can also see a strong coupling of the near-surface airflow with the water motions inside the critical layer. 

The study points out the dynamical role of the airflow motions around the critical layer for relatively young waves, in line with the work of \citet{carpenter2022}. This is in spite of the non-negligible turbulence observed in the airflow within the present simulations, including typical turbulent boundary layer phenomena (ejections and sweeps, detaching spanwise vorticity structures). As such, this study shows a departure from the previously suggested wave age threshold of $15$ suggested by \citet{belcher1998turbulent}, below which only turbulent boundary layer sheltering mechanisms would control the wind energy input into waves. Comparing the two scenarios, numerical results predict a reduction of turbulent stress magnitudes with increasing wave age (decreasing wave slope), which is compensated in parts by wave coherent stresses that transit through the free surface into the thicker critical layer. The pressure contours reveal a significant phase shift and indicate the large influence of the wave slope on the pressure distribution at the surface, respectively, the form drag.

The present numerical approach presents the potential for a future detailed study of the competing contributions of turbulent processes vs.\ mean wave-coherent motions to wave growth \citep{ayet2021dynamical}. 
We have identified the specification of inflow turbulence as an important parameter that can influence the distributions of Reynolds stresses and critical layer heights for younger wave ages, although the basic phase averaged distributions of all quantities are well represented by the simulations. 

The development of this model, with its capability to fully capture coupled atmosphere-ocean dynamics on turbulence-resolving scales, opens up many possibilities for the study of upper ocean and lower atmosphere turbulent processes that have been notoriously difficult to access.  Such processes include energy and momentum fluxes in wave breaking regimes, surface wave phase-resolved Langmuir turbulence, and wind-wave generation mechanisms. Such studies will form the subject of future investigations.

\section*{Acknowledgements}

This paper is a contribution to the project T4 (''Surface Wave-Driven Energy Fluxes at the Air-Sea Interface'') of the Collaborative Research Centre TRR 181 ''Energy Transfers in Atmosphere and Ocean'' funded by the Deutsche Forschungsgemeinschaft (DFG, German Research Foundation) - Grant number 274762653.

The authors gratefully acknowledge the computing time granted by the Resource Allocation Board and provided on the supercomputer Lise and Emmy at NHR@ZIB and NHR@Göttingen as part of the NHR infrastructure. The calculations for this research were conducted with computing resources under the project hhi00037 (''Energy Fluxes at the Air-Sea Interface'').

%
%
\section*{Data Statement}
The CFD simulation data are too large be stored in a public repository. Certain data sets can be made available upon request to Malte Loft (malte.loft@tuhh.de). 

%

\appendix
\section{Blending Functions and Parameters of the IDDES Model}
\label{sec:appendix_blendingFunctions}
The following equations represent the blending functions of the IDDES model used in this study. Equations and parameters agree with a model variant of \citet{gritskevich2012development}.

\begin{align}
    F_1 &= \tanh \left( \left( \min \left[ \max \left[ \frac{\sqrt{k}}{\beta^*  \omega d}, \frac{500 \nu}{d^2 \omega}\right], \frac{2 \rho \omega }{d^2  \frac{\partial k}{\partial x_k} \frac{\partial \omega}{\partial x_k}} \right]\right)^4\right)\\
    F_2 &= \tanh \left(\max\left[\frac{2 \sqrt{k}d}{\omega \beta^*}, \frac{500 \mu }{\rho}\omega d^2\right]^2 \right)\\
    F_\mathrm{d} &= 1-\tanh \left( \left( \frac{c_{\mathrm{dt}1} \mu^\mathrm{t}}{ \rho \kappa^2 d^2 \sqrt{2(|S_{ik}|^2+|W_{ik}|^2)}} \right)^{c_{\mathrm{dt}2}}\right) \\
    \alpha_\mathrm{d} &= 0.25 - d / \Delta \\
    F_\mathrm{b} &= \min \left[ 2 \exp{\left( -9 \alpha_\mathrm{d}^2\right)}, 1 \right] \\
    \tilde{F}_\mathrm{d} &= \max \left[ (1-F_\mathrm{d}), F_\mathrm{b}\right]
\end{align}

\begin{align}
    F_\mathrm{e} &= F_{\mathrm{e}2} \max \left[ F_{\mathrm{e}1} -1,0\right] \\
    F_{\mathrm{e}1} &= \begin{cases}
        2 \exp(-11.09 \alpha_\mathrm{d}^2) &\text{if} \ \alpha_\mathrm{d} \ge 0 \\
        2 \exp(-9.0 \alpha_\mathrm{d}^2) &\text{otherwise}
        \end{cases} \\
    F_{\mathrm{e}2} &= 1 - \max[F_\mathrm{t}, F_\mathrm{l}] \\
    F_\mathrm{t} &=  1-\tanh \left( \left( \frac{c_\mathrm{t}^2 \mu^\mathrm{t}}{ \rho \kappa^2 d^2 \sqrt{2(|S_{ik}|^2+|W_{ik}|^2)}} \right)^3\right) \\
    F_\mathrm{l} &=  1-\tanh \left( \left( \frac{c_\mathrm{l}^2 \mu^\mathrm{t}}{ \rho \kappa^2 d^2 \sqrt{2(|S_{ik}|^2+|W_{ik}|^2)}} \right)^{10}\right)
\end{align}
The following Table \ref{tab:modelConstants} shows all the parameters used. 

\begin{table}[ht]
    \caption{Brief summary of all modeling constants (IDDES)}
    \begin{center}
	\begin{tabular}{ccccc ccccc cccc}
	   \hline
	   &$c_\mu$&$\kappa$&$\beta^*$&$a_1$&$c_\mathrm{DES}$&$\alpha$&$\beta$&$\sigma_\mathrm{k}$&$\sigma_\omega$&$c_\mathrm{w}$&$c_\mathrm{dt}$&$c_\mathrm{l}$&$c_\mathrm{t}$ \\
	   \hline
	   $\varphi_1$&\multirow{2}{*}{$0.09$}&\multirow{2}{*}{$0.41$}&\multirow{2}{*}{$0.09$}&\multirow{2}{*}{$0.31$}&$0.78$&$5/9$&$0.075$&$0.85$&$0.5$&\multirow{2}{*}{$0.15$}&$20$&\multirow{2}{*}{$5.0$}&\multirow{2}{*}{$1.87$} \\
	   $\varphi_2$&&&&&$0.61$&$0.44$&$0.0828$&$1$&$0.856$&&$3$&& \\
	   \hline
        \end{tabular}
    \end{center}
    \label{tab:modelConstants}
\end{table}


\bibliographystyle{plainnat}
\bibliography{library.bib}

\end{document}